\documentclass[aps,floats,showpacs,showkeys,preprintnumbers,nofootinbib]{revtex4}
\pdfoutput=1
\usepackage{amsmath,amssymb}
\usepackage[pdftex,colorlinks=false,urlcolor=blue,linkcolor=blue]{hyperref}

\usepackage{epsfig}

\usepackage{array}
\newcommand{\PreserveBackslash}[1]{\let\temp=\\#1\let\\=\temp}
\newcolumntype{C}[1]{>{\PreserveBackslash\centering}p{#1}}
\newcolumntype{R}[1]{>{\PreserveBackslash\raggedleft}p{#1}}
\newcolumntype{L}[1]{>{\PreserveBackslash\raggedright}p{#1}}

\def\ie{{\it i.e.}}
\def\anti{\overline}

\def\gev{~{\rm GeV}}
\def\tev{~{\rm TeV}}
\def\beq{\begin{equation}}
\def\eeq{\end{\equation}}
\def\bea{\begin{eqnarray}}
\def\eea{\end{eqnarray}}
\def\gam{\gamma}

\def\lam{\lambda}

\def\amu{a_\mu}
\def\damu{\delta \amu}
\def\br{{\rm BR}}

\def\hsm{h_{\rm SM}}

\def\to{\rightarrow}

\def\hpm{H^\pm}

\def\mhpm{m_{\hpm}}
\def\mh{m_h}
\def\mH{m_H}
\def\ma{m_A}
\def\mot{m_{12}}

\def\hsm{h_{\rm SM}}
\def\mhsm{m_{\hsm}}

\def\tanb{\tan\beta}
\def\cotb{\cot\beta}
\def\sinb{\sin\beta}
\def\cosb{\cos\beta}
\def\sina{\sin\alpha}
\def\cosa{\cos\alpha}

\def\beq{\begin{equation}}
\def\eeq{\end{equation}}

\def\eg{{\it e.g.}}
\def\ov{\over}

\def\lsim{\mathrel{\raise.3ex\hbox{$<$\kern-.75em\lower1ex\hbox{$\sim$}}}}
\def\gsim{\mathrel{\raise.3ex\hbox{$>$\kern-.75em\lower1ex\hbox{$\sim$}}}}

\newcommand*{\Ave}[1]{\mathinner{\left\langle{#1}\right\rangle}}
\newcommand{\cp}{\mathcal{CP}}

\def\bit{\begin{itemize}}
\def\eit{\end{itemize}}
\def\bec{\begin{center}}
\def\eec{\end{center}}

\begin{document}
\title{Two-Higgs-Doublet Models and Enhanced Rates for a 125 GeV Higgs}

\author{Aleksandra Drozd$^{1,2}$}
\email[]{aleksandra.drozd@fuw.edu.pl}
\author{Bohdan Grzadkowski$^1$}
\email[]{bohdan.grzadkowski@fuw.edu.pl}
\author{John F.~Gunion$^2$}
\email[]{jfgunion@ucdavis.edu}
\author{Yun~Jiang$^2$}
\email[]{yunjiang@ucdavis.edu}

\affiliation{(1) \,Faculty of Physics, University of Warsaw, 00-681 Warsaw, Poland}
\affiliation{(2) \,Department of Physics, University of California, Davis, CA 95616, USA}

\begin{abstract}
We examine the level of enhancement that can be achieved in the $ZZ$ and $\gam\gam$ channels for a two-Higgs-doublet model Higgs boson (either the light $h$ or the heavy $H$) with mass near $125\gev$ 
after imposing all constraints from LEP data, $B$ physics, precision electroweak data, vacuum stability, unitarity and perturbativity.  
The latter constraints restrict substantially the possibilities for enhancing the $gg\to h \to \gam\gam$ or $gg\to H\to \gam\gam$ signal relative to that 
for the SM Higgs, $\hsm$. Further, we find that a large enhancement of the $gg\to h\to \gamma\gamma$ or $gg\to H\to \gam\gam$  
signal in Type II models is possible only if the  $gg\to h\to ZZ$ or $gg\to H\to ZZ$ mode is even more enhanced, a situation disfavored by current data.  In contrast, in the Type I model one can achieve enhanced rates in the $\gam\gam$ final state for the $h$ while having the $ZZ$ mode at or below the SM rate --- the largest $[gg\to h\to\gam\gam]/[gg\to \hsm \gam\gam]$ ratio  found is of order $\sim 1.3$ when the two Higgs doublet vacuum expectation ratio is  $\tanb =4$ or $20$ and the charged Higgs boson has its minimal LEP-allowed value of $m_{H^\pm}=90\gev$.

\end{abstract}

\pacs{12.60.Fr, 14.80.Ec, 14.80.Fd}
\keywords{Higgs physics, 2-Higgs-Doublet Model, LHC}

\maketitle

\section{Introduction}

The original data from the ATLAS and CMS collaborations~\cite{:2012gk,:2012gu}
provided an essentially $5\sigma$ signal for a Higgs-like resonance with mass of order $123\mbox{--}128\gev$. The updates from Moriond 2013 include those for the $\gam\gam$ channel from ATLAS~\cite{atlasmoriondgamgam} and CMS~\cite{cmsmoriondgamgam}.
The earlier ATLAS and CMS gluon fusion induced rates were significantly enhanced relative to  the Standard Model (SM) prediction. 
The Moriond ATLAS data still shows substantial enhancement for the $\gam\gam$ channel while the CMS MVA analysis finds a roughly SM-like rate in the $\gam\gam$ channel. Here, we consider the extent to which an enhanced $\gam\gam$ rate is possible in various 2HDM models once all relevant theoretical and experimental constraints are imposed

It is known that enhancements with respect to the SM in the $\gam\gam$ channel are generically possible in two-Higgs-doublet models (2HDM) of Type-I and Type-II as explored in \cite{Ferreira:2012my, Ferreira:2011aa, Alves:2012ez, Cheon:2012rh, Chang:2012ve}.  However, these papers do not make clear what level of enhancement is possible after all constraints from B physics and LEP data (B/LEP), precision electroweak data, unitarity and perturbativity are imposed.  
In this paper, we impose all such constraints and determine the maximum possible enhancement. 
We employ a full 1-loop amplitude for Higgs$\to \gamma\gamma$ without neglecting any contributions from 
possible states in the loop.  We examine correlations with other channels. We also consider cases of degenerate scalar 
masses at $\sim 125\gev$~\cite{Gunion:2012gc, Ferreira:2012nv}.

\section{2HDM Models}

The general Higgs sector potential employed is 
\begin{equation}
  \begin{aligned}
    \mathcal{V} = &m_{11}^2\Phi_1^\dagger\Phi_1+m_{22}^2\Phi_2^\dagger\Phi_2
    -\left[m_{12}^2\Phi_1^\dagger\Phi_2+\mathrm{h.c.}\right]
    \\
    &+\frac{1}{2}\lambda_1\left(\Phi_1^\dagger\Phi_1\right)^2
    +\frac{1}{2}\lambda_2\left(\Phi_2^\dagger\Phi_2\right)^2
    +\lambda_3\left(\Phi_1^\dagger\Phi_1\right)\left(\Phi_2^\dagger\Phi_2\right)
    +\lambda_4\left(\Phi_1^\dagger\Phi_2\right)\left(\Phi_2^\dagger\Phi_1\right)
    \\&+\left\{
    \frac{1}{2}\lambda_5\left(\Phi_1^\dagger\Phi_2\right)^2
    +\left[\lambda_6\left(\Phi_1^\dagger\Phi_1\right)
      +\lambda_7\left(\Phi_2^\dagger\Phi_2\right)
      \right]\left(\Phi_1^\dagger\Phi_2\right)
    +\mathrm{h.c.}\right\},
  \end{aligned}
  \label{eq:pot_gen}
\end{equation}
where, to avoid explicit $\cp$ violation in the Higgs sector, all $\lam_i$ and $\mot^2$ are assumed to be real. We choose a basis in which
\[
\Ave{\Phi_1} = \frac{v}{\sqrt{2}}\left(
\begin{array}{c}
0 \\
\cos\beta
\end{array}
\right)\quad\quad
\Ave{\Phi_2} = \frac{v}{\sqrt{2}}\left(
\begin{array}{c}
0 \\
e^{i\xi}\sin\beta 
\end{array}
\right),
\]
where $v=(\sqrt{2}G_F)^{-1/2}\approx246$ GeV. By convention $0\leq \beta \leq \pi/2$ is chosen. For real parameters, the phase $\xi$ could still be non-zero  if the vacuum breaks $\cp$ spontaneously. We avoid parameter choices for which this happens and take $ \xi=0$.  Then, we define
\beq
\Phi_a=\left( 
\begin{array}{c}
\phi_a^+ \cr  (v_a+\rho_a +i \eta_a)/\sqrt 2 
\end{array}
\right)\,\quad a=1,2\,
\eeq
with $v_1=v\cosb$ and $v_2=v\sinb$.  The neutral Goldstone boson is $G^0=\eta_1\cosb+\eta_2\sinb$ while the physical 
pseudoscalar state is 
\beq
A=-\eta_1\sinb+\eta_2\cosb\,.
\eeq
The physical scalars are:
\beq
h=-\rho_1\sina + \rho_2\cosa,\quad H=\rho_1\cosa + \rho_2\sina\,.
\eeq
Without loss of generality, one can assume that
the mixing angle $\alpha$ varies between $-\pi/2$ and $\pi/2$. We choose our independent variables to be $\tanb$ and 
$\sin\alpha$, which are single valued in the allowed ranges. 

We adopt the code 2HDMC~\cite{Eriksson:2010zzb} for numerical calculations. 
All relevant contributions to loop induced processes are taken into account, in particular those with heavy quarks ($t$, $b$ and $c$), $W^\pm$ and $H^\pm$.
A number of different input sets can be used in the 2HDMC context.  We have chosen to use the ``physical basis" in which the inputs are the physical Higgs masses ($\mH,\mh,\ma,\mhpm$), the  
vacuum expectation value ratio ($\tanb$), and the $\cp$-even Higgs mixing angle, $\alpha$, supplemented by $\mot^2$.  The additional parameters  $\lam_6$ and $\lam_7$ are assumed to be zero as a result of a $Z_2$ symmetry being imposed on the dim 4 operators
under which $H_1\to H_1$ and $H_2\to -H_2$. $\mot^2\neq0$ is still allowed as a "soft" breaking of the $Z_2$ symmetry.  With the above inputs, $\lam_{1,2,3,4,5}$ as well as $m_{11}^2$ and $m_{22}^2$ are determined (the latter two via the minimization conditions for a  
minimum of the vacuum)~\cite{Gunion:2002zf}.

In this paper we discuss the Type I and Type II 2HDM models, that are defined by the fermion coupling patterns as 
specified in Table~\ref{fermcoups} --- for more details see \cite{Gunion:1989we}.
\begin{table}[h]
\begin{center}
\begin{tabular}{|c|c|c|c|c|c|c|}
\hline
\ &\multicolumn{3}{c|}{Type I} & \multicolumn{3}{c|}{Type II} \cr
\hline
Higgs & up quarks & down quarks & leptons & up quarks & down quarks & leptons \cr
\hline
 $h$ & $\cosa/ \sinb$ & $\cosa/ \sinb$ & $\cosa/ \sinb$ &  $\cosa/\sinb$ & $-{\sina/\cosb}$  & $-{\sina/ \cosb}$ \cr
\hline
 $H$ & $\sina/ \sinb$ &  $\sina/ \sinb$ &  $\sina/ \sinb$ &  $\sina/ \sinb$ & $\cosa/ \cosb$ & $\cosa/ \cosb$ \cr
\hline
 $A$ & $\cotb$ & $-\cotb$ & $-\cotb$ & $\cotb$ & $\tanb$ & $\tanb$ \cr
\hline 
\end{tabular}
\end{center}
\vspace{-.15in}
\caption{Fermionic couplings $C^{h_i}_{ff}$
normalized to their SM values for the Type I and Type II two-Higgs-doublet models. }
\label{fermcoups}
\end{table}
%
\section{Setup of the Analysis}
The 2HDMC code implements precision electroweak constraints (denoted STU) and limits coming from requiring vacuum stability, 
unitarity and coupling-constant perturbativity (denoted jointly as SUP). 
We note that it is sufficient to consider the SUP constraints at  tree level as usually done in
the literature.  Evolution to higher energies would make these constraints, outlined below,  stronger and would not be appropriate when considering  the 2HDM as an effective low energy theory. In more detail,
the vacuum stability condition requires that the scalar potential be positive in all directions in the limit of growing
field strength~\cite{Deshpande:1977rw}. Tree-level necessary and sufficient conditions for unitarity
are formulated in terms of eigenvalues of the S-matrix in the manner specified in \cite{Ginzburg:2005dt} 
for the most general 2HDM --- the criterion is that the multi-channel Higgs scattering matrix must have a largest eigenvalue below the unitarity limit.
Coupling constant perturbativity is defined as in 2HDMC by the requirement that all self-couplings among the Higgs-boson mass eigenstates be smaller than $4\pi$.
 For the scenarios we consider, this becomes an important constraint on $\lam_1$. 
The SUP constraints are particularly crucial in limiting the level of enhancement  of the $gg\to h\to\gamma\gamma$ 
channel,  which is our main focus. For all our scans, we have supplemented the 2HDMC code by including  the B/LEP constraints.
For the LEP data we adopt upper limits on $\sigma(e^+e^-\to Z\,h/H)$ and $\sigma(e^+e^-\to A\, h/H)$ from 
\cite{Abbiendi:2002qp} and \cite{Abbiendi:2004gn}, respectively.~\footnote{We have modified the subroutine
in 2HDMC that calculates the Higgs boson decays to $\gamma\gamma$ and also the part of the code 
relevant for QCD corrections to the $q\bar q$ final state.}
Regarding $B$ physics, the constraints imposed are those from $\br(B_s\to X_s\gamma)$, $R_b$, 
$\Delta M_{B_s}$, $\epsilon_K$, $\br(B^+\to \tau^+\nu_\tau)$ and $\br(B^+\to D \tau^+\nu_\tau)$. 
The most important implications of these results are to place a lower bound on $\mhpm$ as a function of $\tanb$ 
as shown in Fig.~15 of \cite{Branco:2011iw} in the case of the Type II model and to place a lower bound on 
$\tanb$ as a function of $\mhpm$ as shown in Fig.~18 of \cite{Branco:2011iw} in the case of the Type I model.

While looking for an enhancement of the signal in the $\gamma\gamma$ channel we also computed the extra Higgs-sector
contributions to the anomalous magnetic moment of the muon, $a_\mu=(g_\mu-2)/2$. 
Since the experimentally measured value, $a_\mu = (1165920.80 \pm 0.63) \times 10^{-9}$ ~\cite{Beringer:1900zz},  differs by $\sim 3 \sigma$ from its SM value 
it is important to check 
correlations between $\delta a_\mu\equiv a_\mu- a_\mu^{SM}$ and the signal in the $\gamma\gamma$ channel.
Given the B/LEP, STU and SUP constraints, it turns out that one-loop contributions within the 2HDM are small and negligible, and 
the leading contribution is that known as the Barr-Zee diagram~\cite{Barr:1990vd} which emerges at the two-loop level. 
For completeness we include also sub-leading contributions, see~\cite{Eriksson:2010zzb}. Since the overall
$\sim 3 \sigma$ discrepancy between the experimental and theoretical SM values could still be due to
fluctuations (the world average is based mainly on the E821 result~\cite{Bennett:2004pv} with 
uncertainties  dominated by statistics) or underestimates of the theoretical
uncertainties, we do not use the $a_\mu$ measurement as an experimental constraint on the models we discuss.
However, in tables presented hereafter we do show (in the very last column in units of $10^{-11}$) $\delta a_\mu$, the judgment as to whether $\delta a_\mu$ is acceptable being left to the reader. In fact, for all parameter choices yielding an enhanced Higgs to two-photon rate
the extra contributions to $a_\mu$ are very small and the $a_\mu$ discrepancy is not resolved.


\begin{table}[b]
\begin{center}
\begin{tabular}{|c|C{0.15\textwidth}|C{0.2\textwidth}|C{0.15\textwidth}|C{0.15\textwidth}|C{0.2\textwidth}|}
\hline
 & scenario I & scenario II & scenario III & scenario IV & scenario V \cr
 \hline 
  $m_h$ [GeV] & 125 & \{5,25,45,65,85,105,124.9\} & 125 & 125 &\{5,25,45,65,85,105,124.9\} \cr
 \hline 
   $m_H$ [GeV] & 125+$m_{\rm list}$ & 125 & 125.1 &  125+$m_{\rm list}$ & 125\cr
 \hline 
   $m_A$ [GeV] & $m_{\rm list}$ & $m_{\rm list}$ & $m_{\rm list}$ & 125.1 & 125.1  \cr
 \hline 
  {$m_{H^\pm}$ [GeV]} & \multicolumn{5}{c|} {$\tanb$-dependent minimum value consistent with $B$-physics and other constraints (see caption for details)} \cr
 \hline
 $\tanb$ & \multicolumn{5}{c|}{\{0.5, 0.9, 1.0, 1.2, 1.4, 1.6, 1.8, 2, 3, 4, 5, 7, 10, 15,  20, 30, 40\}} \cr
 \hline
 $\sina$ & \multicolumn{5}{c|}{\{-1, $\ldots$,1\} in steps of 0.1} \cr
 \hline
 $m_{12}^2$ [GeV$^2$]& \multicolumn{5}{c|}{\{$\pm(1000)^2$, $\pm(750)^2$,  $\pm(500)^2$, $\pm(400)^2$, $\pm(300)^2$, $\pm(200)^2$, $\pm(100)^2$, $\pm(50)^2$, $\pm(10)^2$, $\pm(0.1)^2$  \}} \cr
 \hline
\end{tabular}
\end{center}
\vspace{-.15in}
\caption{Range of parameters adopted in the scans. In the table, $m_{\rm list}$ corresponds to the sequence of numbers $m_{\rm list}$=\{0.1,10,50,100,200,300,400,500,750,1000\}. The values of $\mhpm$ are bounded from below
by the constraints from $B$ physics, see Fig.~15 and Fig.~18 of \cite{Branco:2011iw} for the Type II and Type I models,
respectively. We have read off the lower $\mhpm$ bound values at each of the scanned $\tanb$ values from these figures.  Aside from a few preliminary scans, we fix $\mhpm$ at this minimum value while scanning in other parameters.  This is appropriate when searching for the maximum $\gam\gam$ rate since the charged Higgs loop is largest for the smallest possible $\mhpm$.}
\label{parasum}
\end{table}

For an individual Higgs, denoted $h_i$ (where $h_i=h,H,A$ are the choices) we compute the ratio of the $gg$ or 
$WW$-fusion (VBF) induced Higgs cross section times the Higgs branching ratio to a given final state, $X$, 
relative to the corresponding value for the SM Higgs boson as follows:
\begin{eqnarray}
&R^{h_i}_{gg}(X)\equiv (C^{h_i}_{gg})^2\ {\br(h_i\to X)\over \br(\hsm\to X)}, \quad
R^{h_i}_{\rm VBF}(X)\equiv  (C^{h_i}_{WW})^2{\br(h_i\to X)\over\ \br(\hsm\to X)},
\end{eqnarray} 
where $\hsm$ is the SM Higgs boson with $\mhsm=m_{h_i}$ and $C^{h_i}_{gg}, C^{h_i}_{WW}$ are the ratios of the $gg\to h_i$, 
$WW\to h_i$ couplings ($C^A_{WW}$ being zero at tree level) to those for the SM, respectively.  Note that the corresponding ratio for  
$V^*\to V h_i$ ($V=W,Z$) with $h_i\to X$ is equal to $R^{h_i}_{\rm VBF}(X)$, given that kinematic factors cancel out of all these ratios and that these ratios are computed in a self-consistent manner (that is, treating radiative corrections for the SM Higgs boson in the same manner as for the 2HDM Higgs bosons).  When considering cases where 
more than one $h_i$ has mass of $\sim 125\gev$~\cite{Gunion:2012gc}, we sum the different $R^{h_i}$ for the 
production/decay channel of interest. This is justified by the fact that we always choose masses (as indicated in Table~~\ref{parasum}) that are separated by at least $100$~MeV --- in this case interference effects are negligible (given that the Higgs widths are substantially smaller  than 100 MeV in the range of Higgs masses we consider).

We have performed five scans over the parameter space with the range of variation specified in Table~\ref{parasum}. In subsequent tables, if a $\tanb$ value is omitted it means that, for the minimum $\mhpm$ value allowed according to the plots of Ref.~\cite{Branco:2011iw}, no choice for the other parameters could be found for which all the $B$-physics, STU and SUP constraints could be satisfied.

\subsection{\bf \boldmath The $m_h=125\gev$ or $m_H=125\gev$ scenarios:}
%

\begin{figure}[h]
\begin{center}
\includegraphics[width=0.45\textwidth]{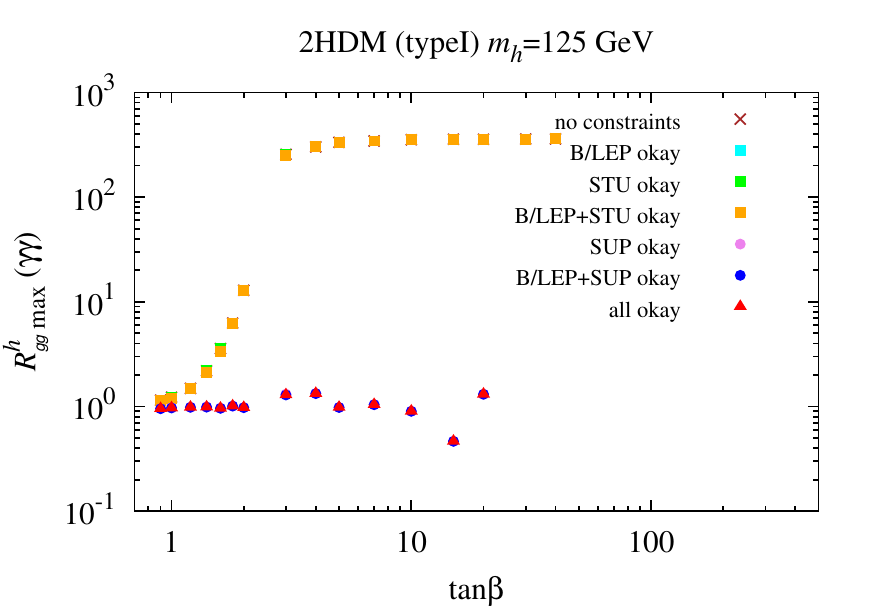}
\includegraphics[width=0.45\textwidth]{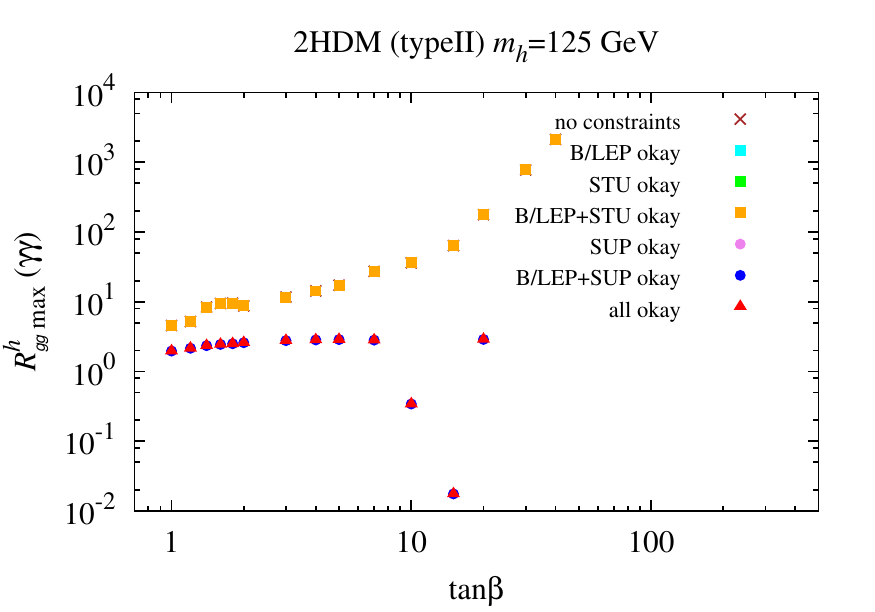}
\includegraphics[width=0.45\textwidth]{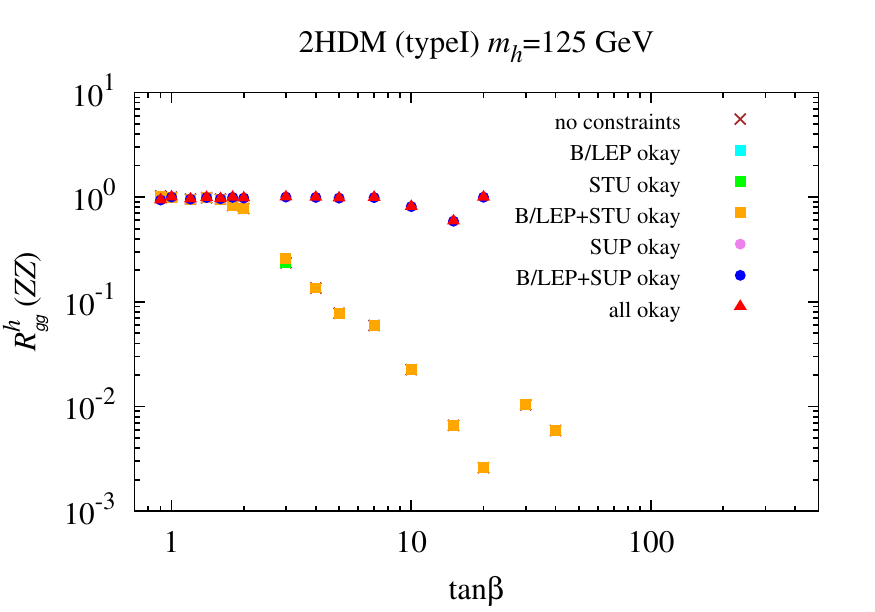}
\includegraphics[width=0.45\textwidth]{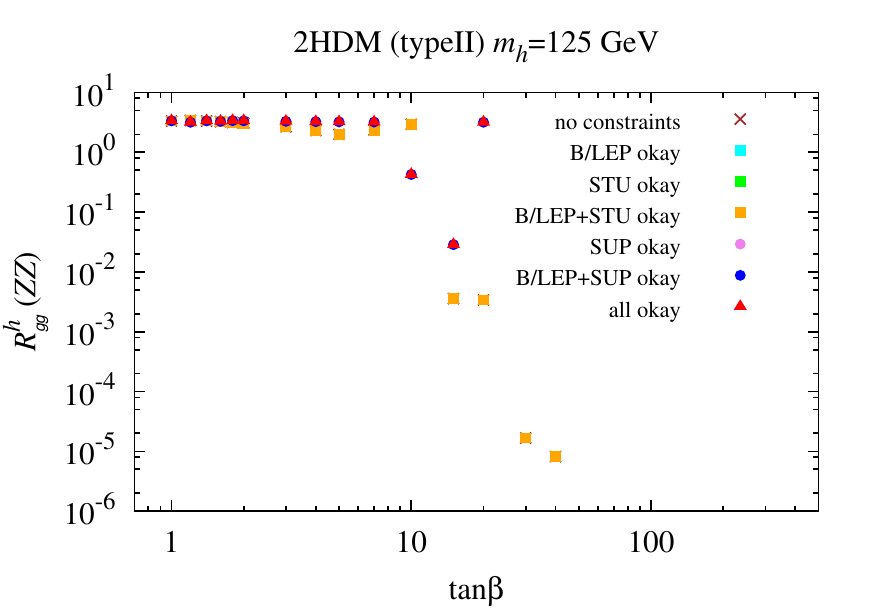}
\includegraphics[width=0.45\textwidth]{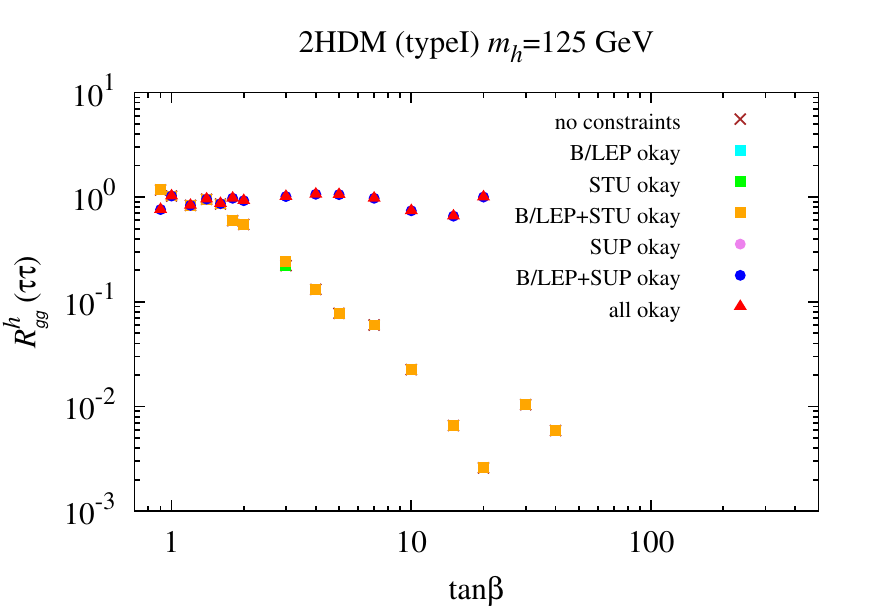}
\includegraphics[width=0.45\textwidth]{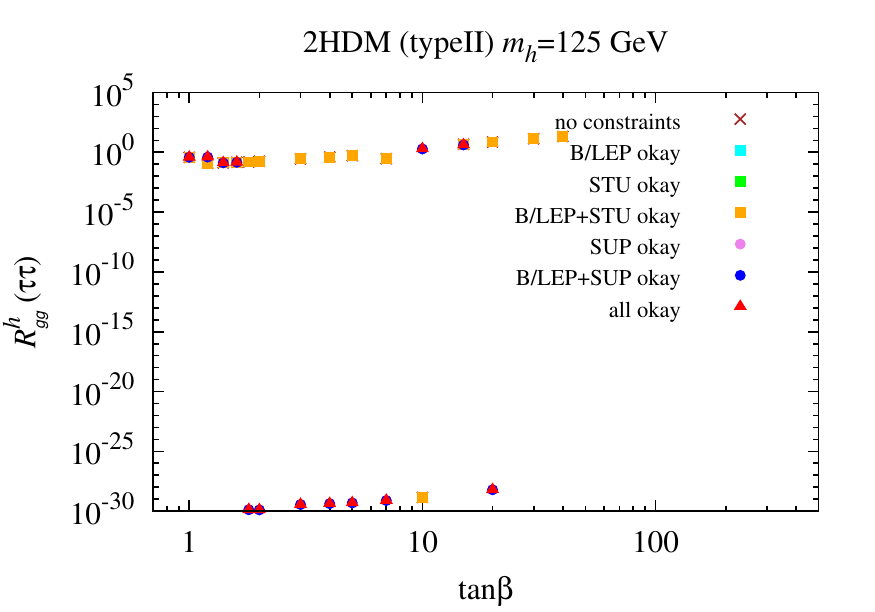}
\end{center}
\vspace{-.25in}
\caption{The top two plots show the maximum $R^h_{gg}(\gam\gam)$ values in the Type I (left) and Type II (right) models for $m_h=125\gev$ as a function of $\tanb$ after imposing various constraints --- see figure legend. { Corresponding $R^h_{gg}(ZZ)$ and $R^h_{gg}(\tau\tau)$ are shown in the middle and lower panels.}
Disappearance of a point after imposing a given constraint set means that the point did not satisfy that set of constraints. 
In the case of boxes and circles, if a given point
satisfies subsequent constraints then the resulting color is chosen according to the color ordering shown in the legend.  This 
same pattern is adopted in the remaining plots. }
\label{honly}
\end{figure}

\begin{table}[b]
\begin{tabular}{|c|ccc|ccc|cccc|c|c|c|}
\hline
$\tan\beta$  &$R^h_{gg\text{max}}(\gam\gam)$  &$R^h_{gg}(ZZ)$  &$R^h_{gg}(\tau\tau)$  &$R^h_{\text{VBF}}(\gam\gam)$  &$R^h_{\text{VBF}}(ZZ)$  &$R^h_{VH}(b\anti{b})$  &$m_H$  &$m_A$  &$m_{H^\pm}$  &$m_{12}$  &$\sin\alpha$  &$\mathcal{A}^h_{H^\pm}/\mathcal{A}$ &$\damu$  \cr  \hline 
 0.9	&0.95	&0.94	&0.76	&1.17	&1.16	&0.94	&875	&750	&900	& 500	&-0.8	&-0.02	&-2.1 \cr  \hline 
 1.0	&0.97	&1.00	&1.02	&0.95	&0.98	&1.00	&875	&750	&850	& 500	&-0.7	&-0.02	&-2.3 \cr  \hline 
 1.2	&0.98	&0.96	&0.83	&1.13	&1.10	&0.96	&625	&750	&612	& 400	&-0.7	&-0.01	&-2.0 \cr  \hline 
 1.4	&0.99	&0.99	&0.96	&1.02	&1.03	&0.99	&525	&750	&460	& 300	&-0.6	&-0.01	&-2.0 \cr  \hline 
 1.6	&0.96	&0.97	&0.87	&1.07	&1.08	&0.97	&625	&400	&360	& 200	&-0.6	&-0.02	&-1.9 \cr  \hline 
 1.8	&1.01	&1.00	&0.98	&1.03	&1.01	&1.00	&425	&400	&285	& 200	&-0.5	&0.00	&-2.0 \cr  \hline 
 2.0	&0.98	&0.98	&0.92	&1.04	&1.04	&0.98	&425	&500	&350	& 200	&-0.5	&-0.01	&-1.8 \cr  \hline 
 3.0	&1.29	&1.00	&1.01	&1.27	&0.99	&1.00	&225	&200	& 92	& 100	&-0.3	&0.12	&-1.8 \cr  \hline 
 4.0	&1.33	&0.99	&1.07	&1.24	&0.93	&0.99	&225	&200	& 90	& 100	&-0.1	&0.14	&-1.7 \cr  \hline 
 5.0	&0.98	&0.98	&1.06	&0.90	&0.91	&0.98	&225	&400	&150	& 100	&-0.0	&0.01	&-1.6 \cr  \hline 
 7.0	&1.04	&0.99	&0.98	&1.06	&1.01	&0.99	&135	&500	& 90	&  50	&-0.2	&0.02	&-1.6 \cr  \hline 
10.0	&0.90	&0.81	&0.74	&0.99	&0.89	&0.81	&175	&500	&150	&  50	&-0.5	&0.04	&-1.5 \cr  \hline 
15.0	&0.46	&0.59	&0.66	&0.41	&0.53	&0.59	&225	&400	&350	&  50	& 0.6	&-0.11	&-1.4 \cr  \hline 
20.0	&1.31	&1.00	&1.00	&1.30	&0.99	&1.00	&225	&200	& 90	&  50	&-0.0	&0.13	&-1.5 \cr  \hline 

\end{tabular}
\vspace{-.1in}
\caption{Table of maximum $R^h_{gg}(\gam\gam)$ values for the Type I 2HDM with $m_h=125\gev$  and associated $R$ values for other initial and/or final states.  The input parameters that give the maximal $R^h_{gg}(\gam\gam)$ value are also tabulated. $\tanb$ values, see Table~\ref{parasum},  for which the full set of constraints cannot be obeyed are omitted.}
\label{tabhIonly}
%
\begin{tabular}{|c|ccc|ccc|cccc|c|c|c|}
\hline
$\tan\beta$  &$R^h_{gg\text{max}}(\gam\gam)$  &$R^h_{gg}(ZZ)$  &$R^h_{gg}(\tau\tau)$  &$R^h_{\text{VBF}}(\gam\gam)$  &$R^h_{\text{VBF}}(ZZ)$  &$R^h_{VH}(b\anti{b})$  &$m_H$  &$m_A$  &$m_{H^\pm}$  &$m_{12}$  &$\sin\alpha$  &$\mathcal{A}^h_{H^\pm}/\mathcal{A}$ &$\damu$  \cr  \hline 
 0.5	&1.56	&2.69	&1.84	&0.52	&0.89	&0.61	&425	&500	&600	& 100	&-0.7	&-0.06	&-0.5 \cr  \hline 
 1.0	&1.97	&3.36	&0.39	&0.65	&1.11	&0.13	&125	&500	&500	& 100	&-0.2	&-0.06	& 0.7 \cr  \hline 
 1.2	&2.16	&3.18	&0.40	&0.95	&1.39	&0.18	&125	&400	&450	& 100	&-0.2	&-0.06	& 0.8 \cr  \hline 
 1.4	&2.35	&3.37	&0.13	&1.07	&1.54	&0.06	&225	&200	&340	& 100	&-0.1	&-0.06	& 1.5 \cr  \hline 
 1.6	&2.45	&3.29	&0.15	&1.30	&1.74	&0.08	&175	&200	&320	& 100	&-0.1	&-0.05	& 1.5 \cr  \hline 
 1.8	&2.51	&3.38	&0.00	&1.31	&1.76	&0.00	&225	&200	&320	& 100	&-0.0	&-0.05	& 1.6 \cr  \hline 
 2.0	&2.59	&3.36	&0.00	&1.48	&1.92	&0.00	&225	&200	&340	& 100	&-0.0	&-0.05	& 1.6 \cr  \hline 
 3.0	&2.78	&3.29	&0.00	&2.01	&2.37	&0.00	&225	&200	&320	& 100	&-0.0	&-0.05	& 1.6 \cr  \hline 
 4.0	&2.84	&3.25	&0.00	&2.24	&2.57	&0.00	&225	&200	&320	& 100	&-0.0	&-0.04	& 1.6 \cr  \hline 
 5.0	&2.87	&3.23	&0.00	&2.37	&2.66	&0.00	&225	&200	&320	& 100	&-0.0	&-0.04	& 1.6 \cr  \hline 
 7.0	&2.83	&3.21	&0.00	&2.42	&2.75	&0.00	&135	&300	&320	&  50	&-0.0	&-0.05	& 0.8 \cr  \hline 
10.0	&0.34	&0.43	&1.89	&0.22	&0.28	&1.23	&325	&200	&320	& 100	& 0.2	&-0.08	& 3.5 \cr  \hline 
15.0	&0.02	&0.03	&4.06	&0.00	&0.01	&0.87	&225	&200	&320	&  50	& 0.6	&-0.14	& 5.3 \cr  \hline 
20.0	&2.89	&3.19	&0.00	&2.57	&2.83	&0.00	&225	&200	&320	&  50	&-0.0	&-0.04	& 2.4 \cr  \hline 

\end{tabular}
\vspace{-.1in}
\caption{Table of maximum $R^h_{gg}(\gam\gam)$ values for the Type II 2HDM with $m_h=125\gev$ and associated $R$ values for other initial and/or final states.  The input parameters that give the maximal $R^h_{gg}(\gam\gam)$ value are also tabulated.  $\tanb$ values, see Table~\ref{parasum}, for which the full set of constraints cannot be obeyed are omitted.
}
\label{tabhIIonly}
\end{table}

Let us begin by discussing the case in which the $h$ has mass $m_h=125\gev$ while scanning over the masses of the other Higgs eigenstates 
(cases where two Higgs are approximaely degenerate are discussed below).
The upper plots of Fig.~\ref{honly} show the maximum value achieved for the ratio $R^h_{gg}(\gam\gam)$ 
as a function of $\tanb$ after scanning over all other input parameters (as specified earlier), in particular $\sina$. 
These maximum values are plotted both prior to imposing any constraints and after imposing various combinations of 
the constraints outlined earlier with point notation as specified in the figure legend. 
We observe that for most values of $\tanb$ the B/LEP and STU precision electroweak constraints, both individually 
and in combination, leave the maximum $R^h_{gg}(\gam\gam)$ unchanged relative to a full scan over all of parameter space.  
In contrast, the SUP constraints greatly reduce the maximum value of $R^h_{gg}(\gam\gam)$ that can be achieved and 
that value is left unchanged when B/LEP and STU constraints are imposed in addition.  
Remarkably, in the Type I model maximum $R^h_{gg}(\gam\gam)$ 
values much above $1.3$ are not possible, with values close to $1$ being more typical for most $\tanb$ values. In contrast,  maximum $R^h_{gg}(\gam\gam) $ values in the range of $2-3$ 
are possible for $2 \leq\tanb\leq 7$ and $\tanb=20$ in the Type II model.  In Fig.~\ref{honly} we also show the values of $R^h_{gg}(ZZ)$ and $R^h_{gg}(\tau\tau)$ (middle and bottom plots, respectively) found for those parameter choices giving the maximum $R^h_{gg}(\gam\gam) $ values appearing in the upper plots.

\begin{figure}[t]
\begin{center}
\includegraphics[width=0.45\textwidth]{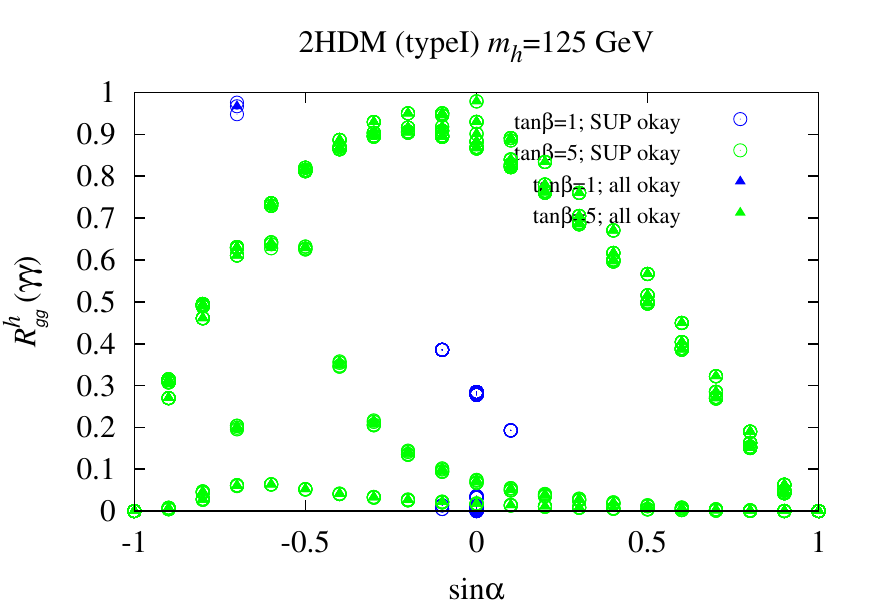}
\includegraphics[width=0.45\textwidth]{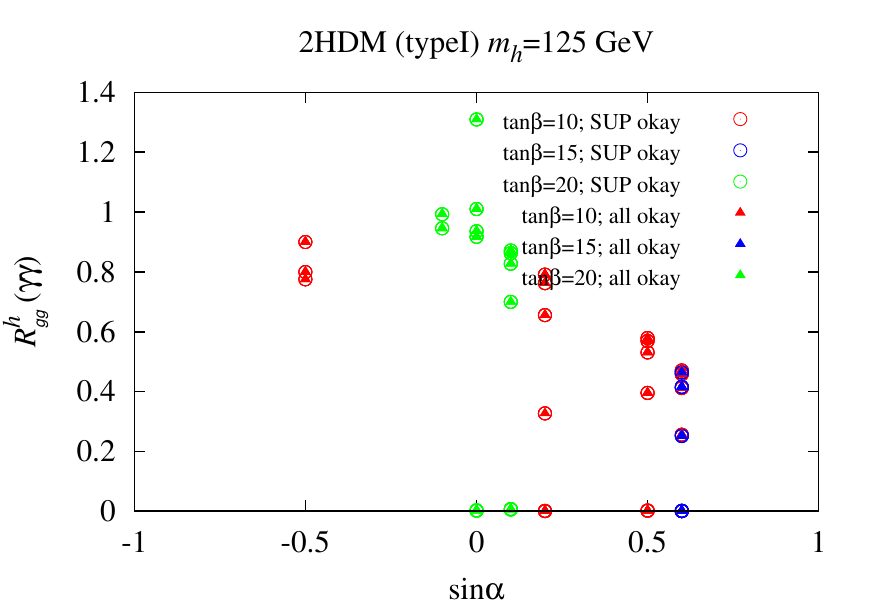}
\includegraphics[width=0.45\textwidth]{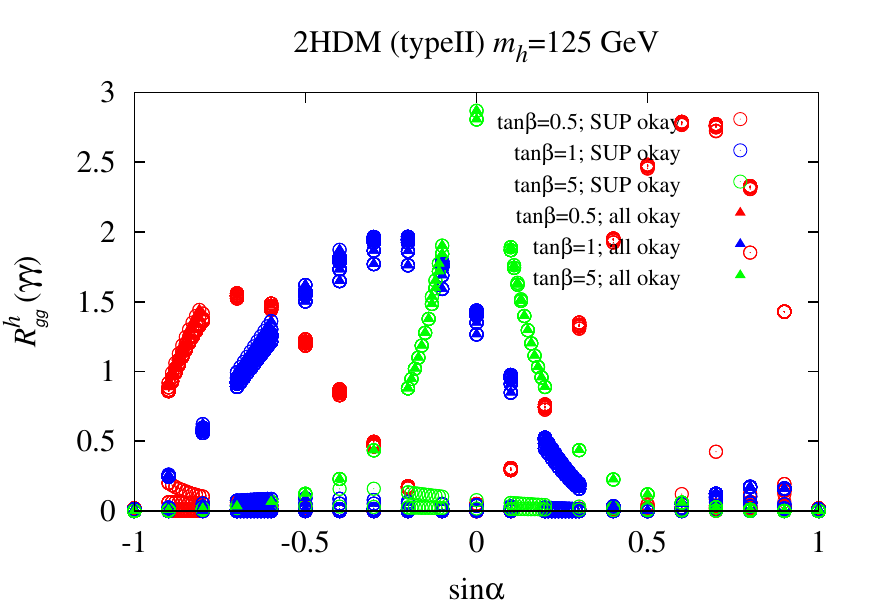}
\includegraphics[width=0.45\textwidth]{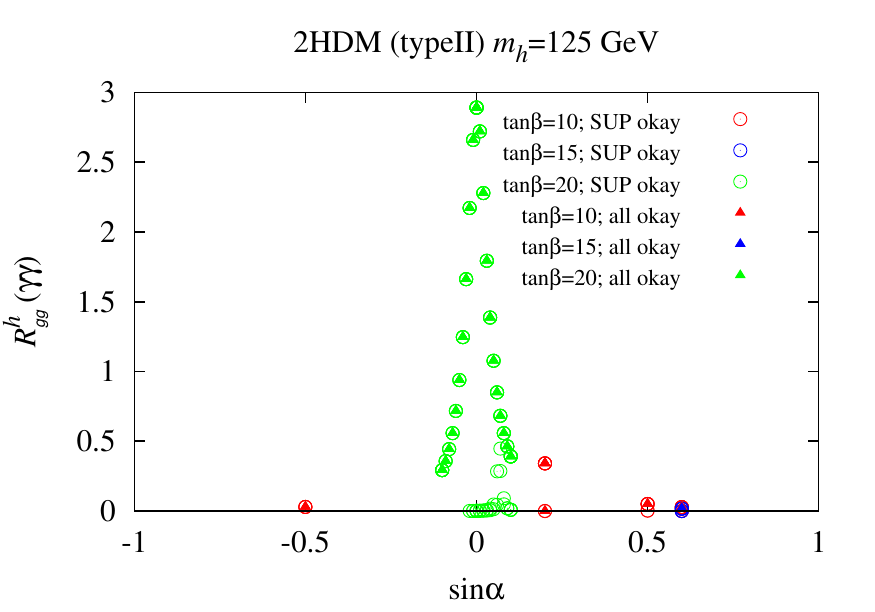}
\vspace{-.15in}
\caption{$R^h_{gg}(\gam\gam)$ is plotted for $m_h=125\gev$ as a function of $\sina$ for a sequence of $\tanb$ values.  
Different constraint combinations are considered and the different curves of a given type correspond to a variety 
of other input parameters.  The upper plots are for the Type I model and the lower plots are for the Type II model.  
Different colors indicate different $\tanb$ values. $\tanb$ values, see Table~\ref{parasum}, for which the full set of constraints cannot be obeyed are omitted.}
\label{honlysinaplot}
\end{center}
\end{figure}

One can get a feeling for how the different constraints impact $R^h_{gg}(\gam\gam)$ by plotting this quantity as 
a function of $\sina$ at fixed $\tanb$ for different constraint combinations and a selection of different other 
input parameters.
As shown in Fig.~\ref{honlysinaplot}, $R^h_{gg}(\gam\gam)$ typically has a maximum as $\sina$ is varied but the 
height of this maximum depends very much on the constraints imposed as there is also variation with the other input parameters.

Tables~\ref{tabhIonly} and \ref{tabhIIonly} display the full set of input parameters corresponding to the maximal $R^h_{gg}(\gam\gam)$ values at each $\tanb$ for models of Type I and Type II, respectively.
It is important to notice that in the Type II model, the value of $R^h_{gg}(ZZ)$ corresponding to the parameters 
that maximize $R^h_{gg}(\gam\gam)$ is typically large, $\sim 3$. In fact, as discussed shortly, $R^h_{gg}(ZZ)> R^h_{gg}(\gam\gam)$ whenever $R^h_{gg}(\gam\gam)$ is even modestly enhanced. The current experimental situation is confused. 
The Moriond 2013 ATLAS data~\cite{atlasmoriondgamgam} shows  central values of $R^h_{gg}(\gam\gam)\sim 1.6$ and $R^h_{gg}(ZZ)\sim 1.5$. In the Type II model case, the former would imply $R^h_{gg}(ZZ)> 2$, somewhat inconsistent with the observed central value.  However, the data uncertainties are significant and so it is too early to conclude that the Type II model cannot describe the ATLAS data. The Moriond  2013 CMS data has central values of  $R^h_{gg}(\gam\gam)<1$ and $R^h_{gg}(ZZ)\sim 1$, a situation completely consistent with the Type II model predictions.

As an aside, we note that  $R^h_{gg}(\gam\gam)/R^h_{gg}(ZZ)\gsim1$ when $R^h_{gg}(\gam\gam)>1$  is fairly typical of the MSSM model (which has a Type II Higgs sector), especially with full or partial GUT scale unification for the soft-SUSY-breaking parameters, see for example \cite{Brummer:2012ns}. In such scenarios the primary modification to the $\gam\gam$ rate relative to the SM is due to the light stop loop contribution to the $h\gam\gam$ coupling (which enters with the same sign as the $W$ loop and has a color factor enhancement)  which enhances $\br(h\to \gam\gam)$.  Note that the stop loop contribution to the $hgg$ production coupling is the same for both the $ZZ$ and $\gam\gam$ final states.  In the absence of GUT scale unification, there are many other potentially significant loops contributing to an increase in the $h\gam\gam$ coupling, the most important being the light chargino loop and the light stau loop, as studied for example in~\cite{Carena:2012gp}.

\begin{table}[h]
\begin{tabular}{|c|ccc|ccc|cccc|c|c|c|}
\hline
$\tan\beta$  &$R^H_{gg\text{max}}(\gam\gam)$  &$R^H_{gg}(ZZ)$  &$R^H_{gg}(\tau\tau)$  &$R^H_{\text{VBF}}(\gam\gam)$  &$R^H_{\text{VBF}}(ZZ)$  &$R^H_{VH}(b\anti{b})$  &$m_h$  &$m_A$  &$m_{H^\pm}$  &$m_{12}$  &$\sin\alpha$  &$\mathcal{A}^H_{H^\pm}/\mathcal{A}$ &$\damu$  \cr  \hline 
 1.4	&0.91	&0.99	&0.96	&0.94	&1.03	&0.99	&125	&500	&460	& 100	& 0.8	&0.00	&-2.7 \cr  \hline 
 1.6	&0.89	&0.97	&0.87	&1.00	&1.08	&0.97	&125	&400	&360	&  50	& 0.8	&0.00	&-2.5 \cr  \hline 
 1.8	&0.89	&1.01	&1.08	&0.84	&0.95	&1.01	&125	&400	&350	&  50	& 0.9	&-0.06	&-2.3 \cr  \hline 
 2.0	&0.90	&1.00	&1.02	&0.89	&0.99	&1.00	&125	&400	&350	&  50	& 0.9	&-0.05	&-2.1 \cr  \hline 
 3.0	&0.89	&0.96	&0.88	&0.97	&1.05	&0.96	&125	&400	&350	&  50	& 0.9	&-0.05	&-1.8 \cr  \hline 
 4.0	&0.89	&0.97	&1.09	&0.79	&0.86	&0.97	&105	&500	& 90	&  50	& 1.0	&-0.03	&-1.7 \cr  \hline 
 5.0	&0.93	&0.98	&1.06	&0.86	&0.90	&0.98	&125	&500	& 90	&  50	& 1.0	&-0.01	&-1.6 \cr  \hline 
 7.0	&0.88	&0.99	&1.03	&0.85	&0.95	&0.99	& 65	&400	&350	&  10	& 1.0	&-0.05	&-1.6 \cr  \hline 
10.0	&0.89	&1.00	&1.02	&0.87	&0.98	&1.00	& 45	&400	&350	&   0	& 1.0	&-0.05	&-1.6 \cr  \hline 
15.0	&0.90	&1.00	&1.01	&0.89	&0.99	&1.00	&  5	&400	&350	&   0	&-1.0	&-0.05	&-1.6 \cr  \hline 
20.0	&0.90	&1.00	&1.00	&0.89	&0.99	&1.00	& 25	&400	&350	&   0	&-1.0	&-0.05	&-1.5 \cr  \hline 
30.0	&0.90	&1.00	&1.00	&0.90	&1.00	&1.00	&  5	&400	&350	&   0	&-1.0	&-0.05	&-1.5 \cr  \hline 
40.0	&0.90	&1.00	&1.00	&0.90	&1.00	&1.00	&  5	&400	&350	&   0	&-1.0	&-0.05	&-1.5 \cr  \hline 

\end{tabular}
\vspace{-.1in}
\caption{Table of maximum $R^H_{gg}(\gam\gam)$ values for the Type I 2HDM with $m_H=125\gev$  and associated $R$ values for other initial and/or final states.  The input parameters that give the maximal $R^H_{gg}(\gam\gam)$ value are also tabulated. $\tanb$ values, see Table~\ref{parasum}, for which the full set of constraints cannot be obeyed are omitted.}
\label{tabHIonly}
%
%
%
\begin{tabular}{|c|ccc|ccc|cccc|c|c|c|}
\hline
$\tan\beta$  &$R^H_{gg\text{max}}(\gam\gam)$  &$R^H_{gg}(ZZ)$  &$R^H_{gg}(\tau\tau)$  &$R^H_{\text{VBF}}(\gam\gam)$  &$R^H_{\text{VBF}}(ZZ)$  &$R^H_{VH}(b\anti{b})$  &$m_h$  &$m_A$  &$m_{H^\pm}$  &$m_{12}$  &$\sin\alpha$  &$\mathcal{A}^H_{H^\pm}/\mathcal{A}$ &$\damu$  \cr  \hline 
 1.0	&1.99	&3.24	&0.52	&0.71	&1.16	&0.19	&125	&500	&500	& 100	& 1.0	&-0.06	& 0.7 \cr  \hline 
 1.2	&1.90	&3.34	&0.00	&0.59	&1.04	&0.00	&125	&400	&450	& 100	& 1.0	&-0.06	& 0.8 \cr  \hline 
 1.4	&2.20	&3.38	&0.00	&0.86	&1.32	&0.00	&125	&300	&340	& 100	& 1.0	&-0.06	& 1.1 \cr  \hline 
 1.6	&2.36	&3.39	&0.00	&1.09	&1.56	&0.00	&125	&300	&320	&  50	& 1.0	&-0.06	& 1.1 \cr  \hline 
 1.8	&2.48	&3.38	&0.00	&1.29	&1.76	&0.00	&125	&300	&320	&  50	& 1.0	&-0.06	& 1.1 \cr  \hline 
 2.0	&2.56	&3.36	&0.00	&1.46	&1.92	&0.00	&125	&300	&340	&  50	& 1.0	&-0.06	& 1.1 \cr  \hline 
 3.0	&2.73	&3.29	&0.00	&1.97	&2.37	&0.00	&125	&300	&320	&  50	& 1.0	&-0.05	& 1.0 \cr  \hline 
 4.0	&2.78	&3.25	&0.00	&2.20	&2.57	&0.00	&125	&300	&320	&  50	&-1.0	&-0.05	& 1.0 \cr  \hline 
 5.0	&2.81	&3.23	&0.00	&2.32	&2.66	&0.00	&125	&300	&320	&  50	&-1.0	&-0.05	& 0.9 \cr  \hline 
 7.0	&2.80	&3.21	&0.00	&2.40	&2.75	&0.00	& 65	&300	&320	&  10	&-1.0	&-0.06	&-0.0 \cr  \hline 
10.0	&2.81	&3.20	&0.00	&2.46	&2.79	&0.00	& 45	&300	&320	&   0	&-1.0	&-0.06	&-2.8 \cr  \hline 
15.0	&2.82	&3.19	&0.00	&2.49	&2.82	&0.00	& 25	&300	&320	&   0	&-1.0	&-0.05	&-16.9 \cr  \hline 
20.0	&2.82	&3.19	&0.00	&2.50	&2.83	&0.00	& 25	&300	&320	&   0	&-1.0	&-0.05	&-30.8 \cr  \hline 
30.0	&2.82	&3.19	&0.00	&2.51	&2.84	&0.00	&  5	&300	&320	&   0	&-1.0	&-0.05	&-38.7 \cr  \hline 
40.0	&2.82	&3.19	&0.00	&2.51	&2.84	&0.00	&  5	&300	&320	&   0	&-1.0	&-0.05	&-69.7 \cr  \hline 

\end{tabular}
\vspace{-.1in}
\caption{Table of maximum $R^H_{gg}(\gam\gam)$ values for the Type II 2HDM with $m_H=125\gev$ and associated $R$ values for other initial and/or final states.  The input parameters that give the maximal $R^H_{gg}(\gam\gam)$ value are also tabulated. $\tanb$ values. see Table~\ref{parasum}, for which the full set of constraints cannot be obeyed are omitted.
}
\label{tabHIIonly}
\end{table}

Corresponding results for the $H$ are presented for the Type I and Type II models in Tables~\ref{tabHIonly} and \ref{tabHIIonly}, respectively.  In the case of the Type I model, an enhanced gluon fusion rate in the $\gam\gam$ final state does not seem to be possible after imposing the SUP constraints, whereas maximal enhancements of order $R^H_{gg}(\gam\gam)\sim 2.8$ are quite typical for the Type II model, albeit with even larger $R^H_{gg}(ZZ)$.  Again, in the case of the Type II model $R^H_{gg}(\gam\gam)/R^H_{gg}(ZZ)<1$ applies more generally whenever $R^H_{gg}(\gam\gam)$ is significantly enhanced.

\begin{figure}[h]
\begin{center}
\includegraphics[width=0.45\textwidth]{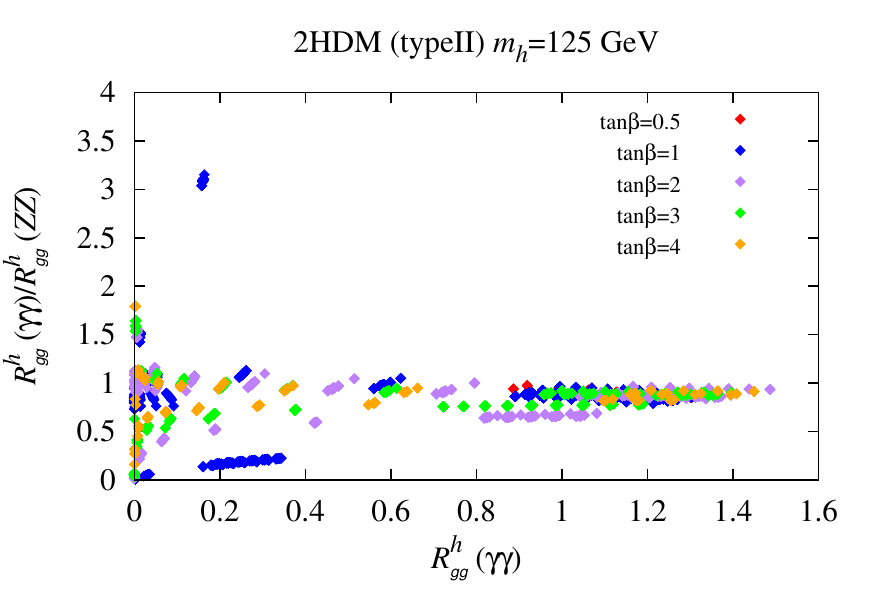}
\includegraphics[width=0.45\textwidth]{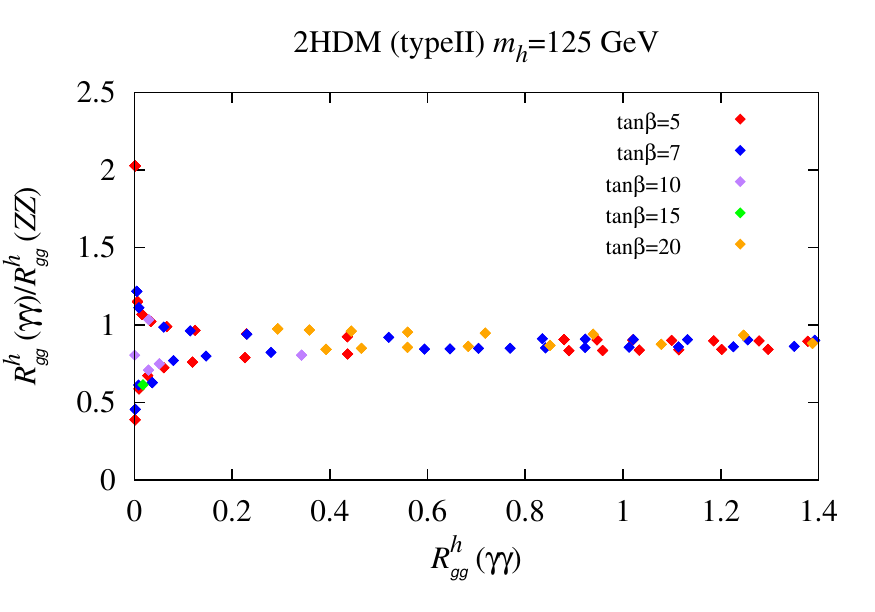}
\includegraphics[width=0.45\textwidth]{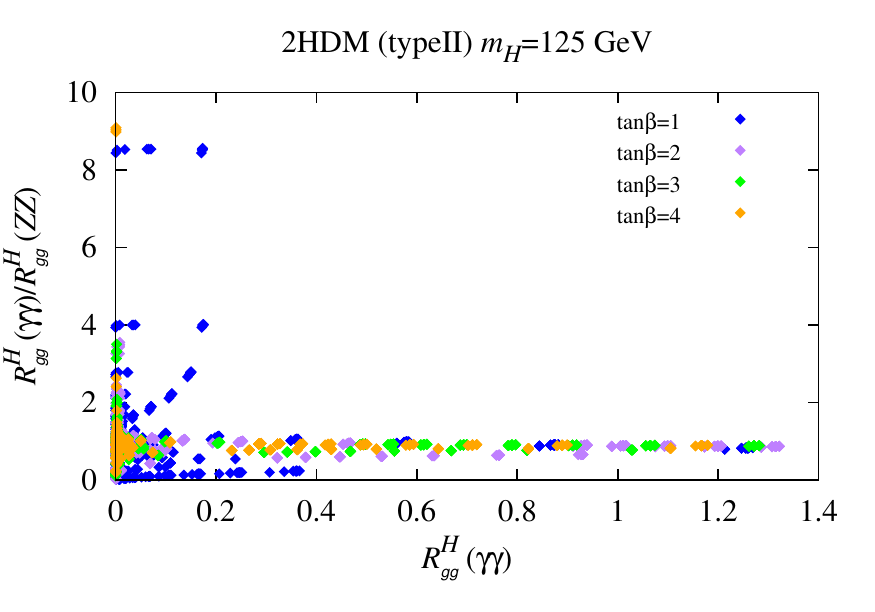}
\includegraphics[width=0.45\textwidth]{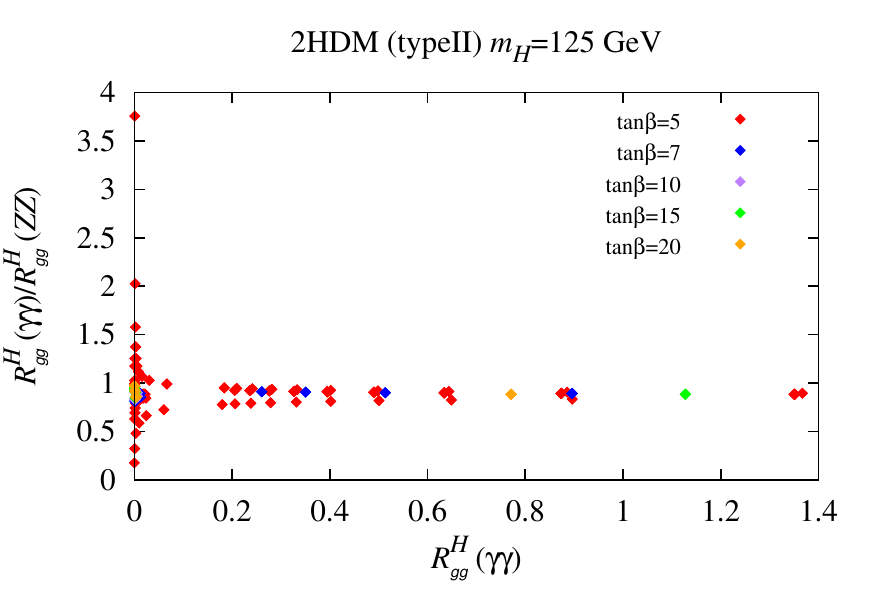}
\end{center}
\caption{For the $h$ (top) and $H$ (bottom) we plot the gluon fusion induced  $\gam\gam/ZZ$ ratio as a function of $R_{gg}(\gam\gam)$ for the Type II 2HDM. \label{IIcorr}}
\end{figure}

\begin{figure}[t]
\begin{center}
\includegraphics[width=0.42\textwidth]{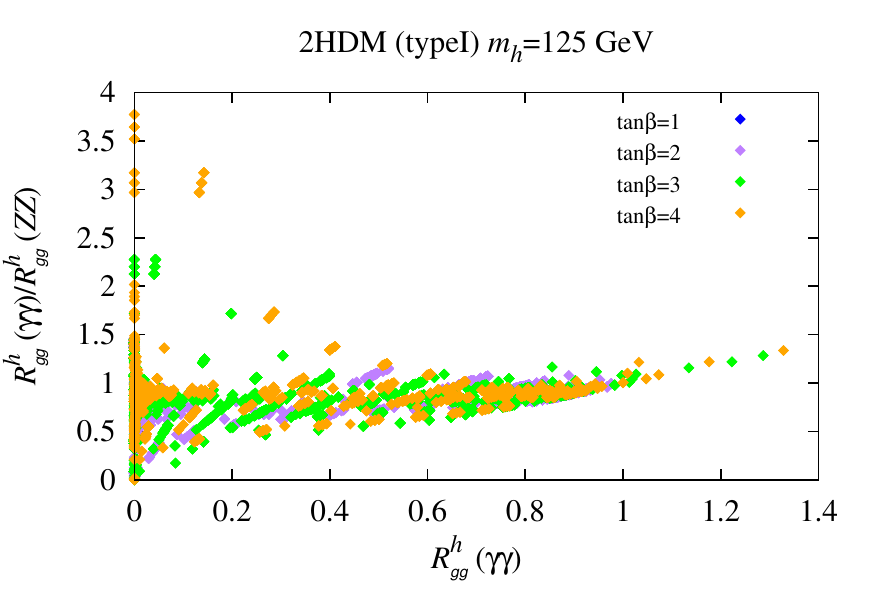}
\includegraphics[width=0.42\textwidth]{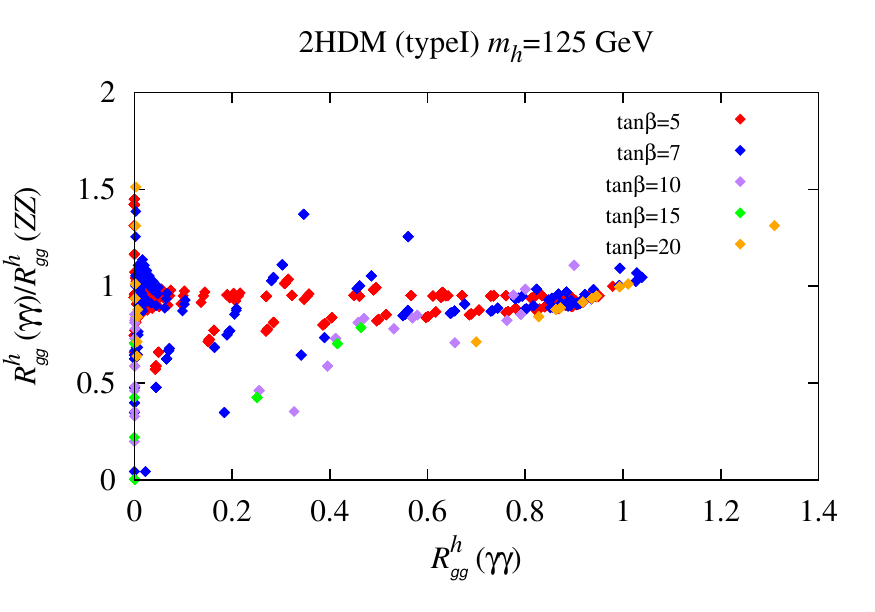}
\includegraphics[width=0.42\textwidth]{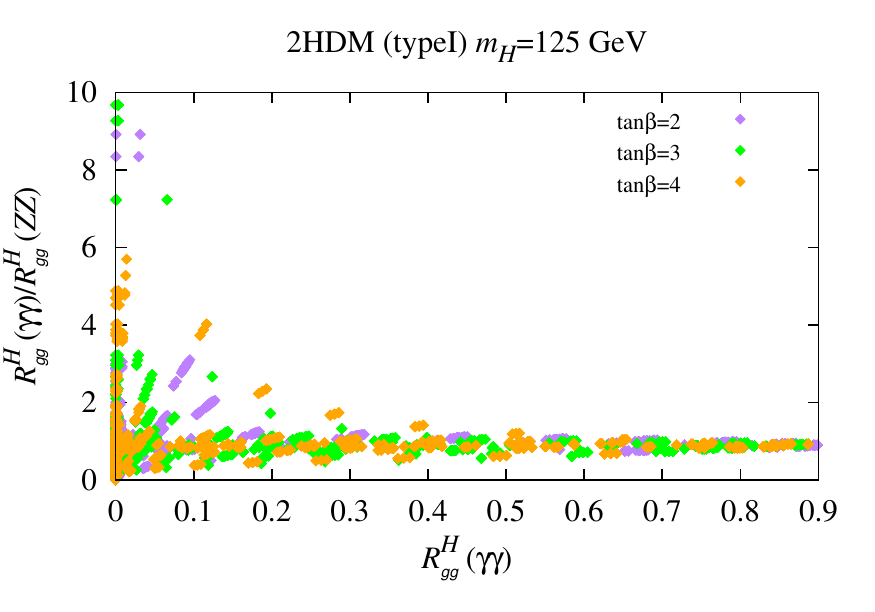}
\includegraphics[width=0.42\textwidth]{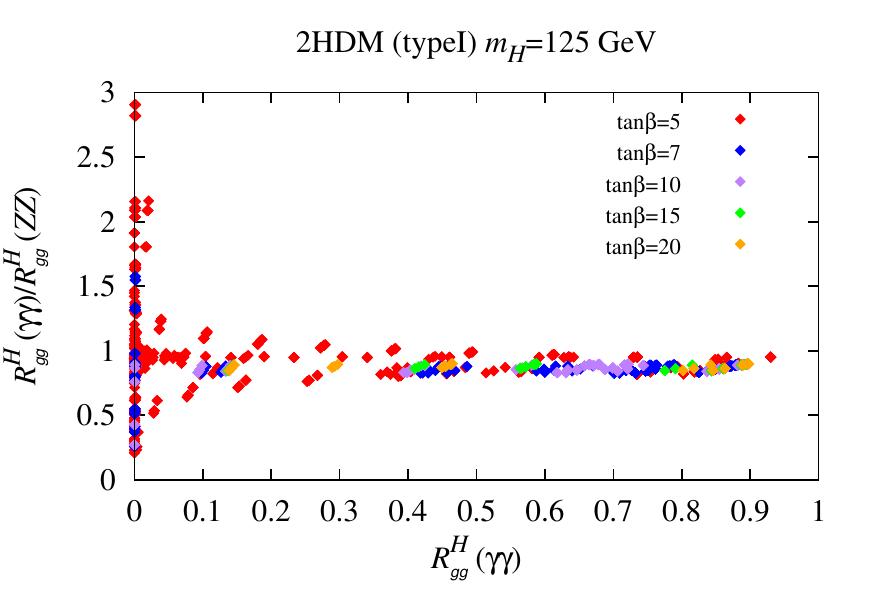}
\end{center}
\caption{For the $h$ (top) and $H$ (bottom) we plot the gluon fusion induced  $\gam\gam/ZZ$ ratio as a function of $R_{gg}(\gam\gam)$ for the Type I 2HDMl. \label{Icorr}}
\end{figure}

That  an enhanced $\gam\gam$ rate, \eg\ $R^{h,H}_{gg}(\gam\gam)>1.2$, leads to $R^{h,H}_{gg}(\gam\gam)/R^{h,H}_{gg}(ZZ)<1$ in Type II models is illustrated by the plots of Fig.~\ref{IIcorr}.  We again emphasize that this is to be contrasted with the Type I model for which  $R^{h}_{gg}(\gam\gam) >1.2$ implies  $R^{h}_{gg}(\gam\gam)/R^{h,H}_{gg}(ZZ)>1$,  see Fig.~\ref{Icorr}, in better agreement with current data.  (For the Type I model, $R^{H}_{gg}(\gam\gam) >1$ is not possible after imposing the SUP constraints.)

It is interesting to understand the mechanism behind the enhancement of $R^{h,H}_{gg}(ZZ)$ that seems to be
an inevitable result within the Type II model if $R^{h,H}_{gg}(\gam\gam)$ is large. Let us define $r$ as the ratio of $\gamma\gamma$ over $ZZ$ production rates for a scalar $s$ (either $h$ or $H$). Then it is easy to see that
\beq
r_s\equiv{R^s_{gg}(\gam\gam)\ov R^s_{gg}(ZZ) }={\Gamma(s \to \gam\gam)/\Gamma(h_{sm} \to \gam\gam) \ov \Gamma(s\to ZZ)/\Gamma(h_{sm}\to ZZ)}\,.
\label{rdef}
\eeq
For the decay mode $s \to ZZ^*$, the tree level amplitude is present and dominant so that the denominator simply reduces to $(C^s_{ZZ})^2$.
For the decay mode $s \to \gam\gam$, there is no tree level contribution --- the $s\gam\gam$ coupling first arises at the one-loop level with the $t$-loop, $W$-loop and $H^\pm$-loop being the important contributions. As a result, the numerator can be written as
\beq
{\Gamma(s \to \gam\gam) \ov \Gamma(h_{sm} \to \gam\gam)}=\left({C^s_{WW} \mathcal{A}^{SM}_W - C^s_{t\bar t} \mathcal{A}^{SM}_t + \mathcal{A}_{H^\pm} \ov \mathcal{A}^{SM}_W-\mathcal{A}^{SM}_t}\right)^2
\eeq
where $C^s_{t\bar t}$ and $C^s_{WW}$ are the $st\bar t$ and $s WW$ couplings normalized to those of the $\hsm$, while $\mathcal{A}^{SM}_W$ and $\mathcal{A}^{SM}_t$ are the $W$-loop and $t$-loop amplitudes, respectively, for the $\hsm$. Finally, $\mathcal{A}_{H^\pm}$ is the $H^\pm$-loop amplitude in the 2HDM; since it is very small in the Type II model, it can be neglected. Thus, 
\beq
r_s \simeq {(C^s_{WW})^2 \ov (C^s_{ZZ})^2 }\left({\mathcal{A}^{SM}_W-{C^s_{t\bar t} \ov C^s_{WW}} \mathcal{A}^{SM}_t \ov \mathcal{A}^{SM}_W-\mathcal{A}^{SM}_t}\right)^2 = \left({\mathcal{A}^{SM}_W-{C^s_{t\bar t} \ov C^s_{WW}} \mathcal{A}^{SM}_t \ov \mathcal{A}^{SM}_W-\mathcal{A}^{SM}_t}\right)^2
\eeq
where $C^s_{ZZ}=C^s_{WW}$ in any doublets+singlets models. Note that when the $t$-loop contribution is negligible then $r_s\to 1$. 
It is easy to see that  $r_s<1$ if the following inequality is satisfied
\beq
1<{C^s_{t\bar t} \ov C^s_{WW}} < 2 {\mathcal{A}^{SM}_W \ov \mathcal{A}^{SM}_t} - 1 
\label{r1}
\eeq
When $C^s_{t\bar t} / C^s_{WW}$ is outside of the above interval then $r_s>1$.
If $s$ is the lighter scalar $h$ then  $C^s_{t\bar t} / C^s_{WW}=\cos\alpha / [\sin\beta\sin(\beta-\alpha)]$ implying
$r_h<1$ when
\beq
1< {\cos\alpha \ov \sin\beta\sin(\beta-\alpha)} < 2 {\mathcal{A}^{SM}_W \ov \mathcal{A}^{SM}_t} - 1 \simeq 9\,,
\label{caseh}
\eeq
while for $s=H$, $C^s_{t\bar t} / C^s_{WW}=\sin\alpha / [\sin\beta\cos(\beta-\alpha)]$  and we obtain $r_H<1$ for
\beq
1< {\sin\alpha \ov \sin\beta\cos(\beta-\alpha)} < 2 {\mathcal{A}^{SM}_W \ov \mathcal{A}^{SM}_t} - 1 \simeq 9.
\label{caseH}
\eeq
In the case $s=h$, $R^h_{gg}(\gam\gam)$ is maximized by suppressing the $h$ total width, which corresponds to chosing
$\alpha$ so as to minimize the $hb\anti b$ coupling,  \ie\  $\alpha \sim 0$, resulting in
$ C^h_{t\bar t} / C^h_{WW} \sim  1/\sin^2\beta > 1$  (and $<5$ for $\tanb>0.5$). Consequently $r_h<1$, as observed in Table~\ref{tabhIIonly}.
The argument is similar in the case of the $H$: this time the $Hb\anti b$ coupling is chosen to be small (equivalent to  $\alpha\sim \pm \pi/2$) in order  to minimize the $H$ total width  and therefore maximize $R^H_{gg}(\gam\gam)$, 
with the result that once again $C^H_{t\bar t} / C^H_{WW} \sim 1/\sin^2\beta > 1$, yielding $r_H<1$. These analytic results explain why large $R^s_{gg}(\gam\gam)$ is correlated with even larger $R^s_{gg}(ZZ)$ in Type II 2HDMs.

\begin{figure}[h]
\begin{center}
\includegraphics[width=0.45\textwidth]{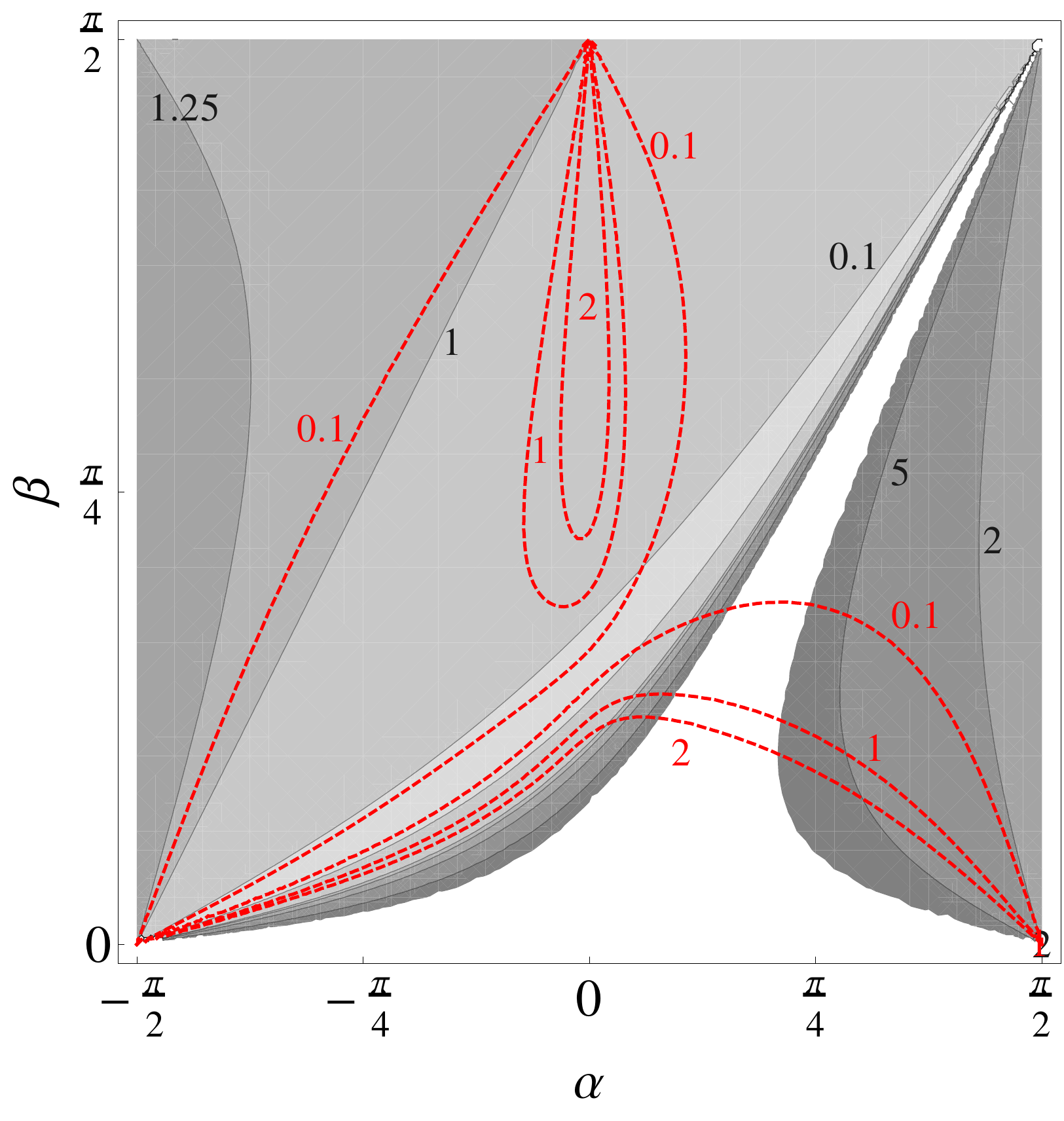}
\includegraphics[width=0.45\textwidth]{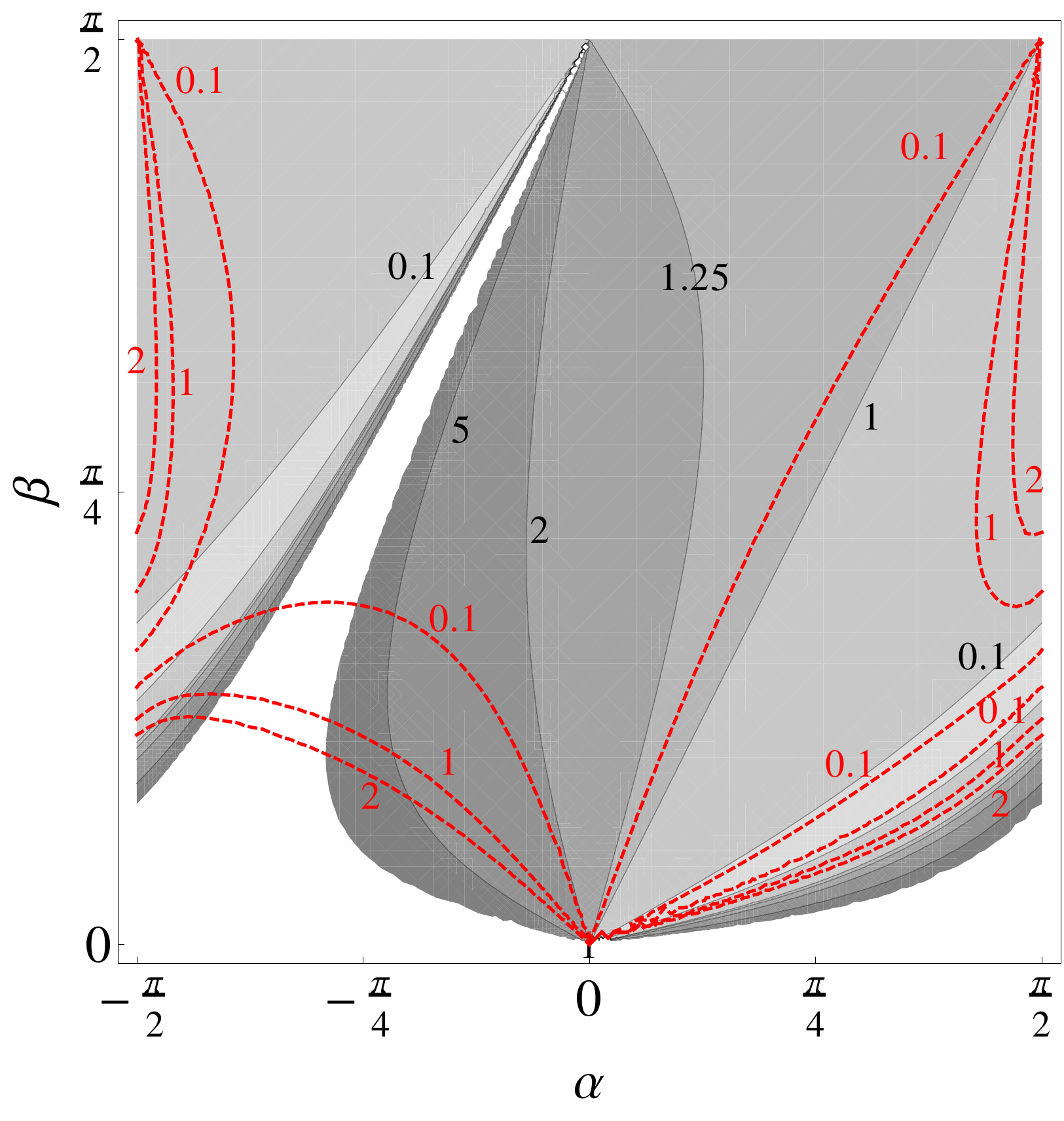}
\vspace{-.15in}
\caption{In the left panel we show contour plots (with shadowing) in the $(\beta,\alpha)$ space for $r_h$ with 
superimposed red contours for $R^h_{gg}(\gam\gam)$. The right panel show similarly $r_H$ with $R^H_{gg}(\gam\gam)$. 
Red numbers give constant values of $R^h_{gg}(\gam\gam)$ ($R^H_{gg}(\gam\gam)$) while black ones show constant 
values of $r_h$ ($r_H$). The white region correspond to $r_s > 10.75$. }
\label{rcontours}
\end{center}
\end{figure}

In Fig.~\ref{rcontours} we plot contours of $r_s$ and of $R^s_{gg}(\gam\gam)$ 
in the Type II model. It is seen from the left panel that if $\tanb$ is large then only
small $\alpha$'s will maximize $R^s_{gg}(\gam\gam)$. And, in that region, 
$r_h$ is always less than $1$.  
Note that the $R^s_{gg}(\gam\gam)> 1$ region shrinks for large $\tanb$, so the the values of $\alpha$ preferred for large $R^h_{gg}(\gam\gam)$    
converge to $0$ when $\beta \to \pi/2$. For the case of
$s=H$,  the right panel shows that when $\tanb$ is large then only vertical bands of $\alpha$ corresponding to values 
close to $\pm \pi/2$ are allowed if $R^s_{gg}(\gam\gam)> 1$. 
From the plots, we see that $R^s_{gg}(\gam\gam)> 1$ could be consistent with $r_s>1$ only if $\tanb\lsim 1$,
which explains the pattern observed in Tables~\ref{tabhIIonly} and \ref{tabHIIonly} and Fig.\ref{IIcorr}. Note, however,
that small $\tanb$ is disfavored by B-physics as it enhances the $H^+\bar{t}b$ coupling too much, see for example~\cite{Branco:2011iw}.

Once again, we emphasize that a  substantial enhancement of the $\gam\gam$ rate {\it is} possible for the $h$ in Type I models without enhancing the $ZZ$ rate.   In particular, from Table~\ref{tabhIonly} we see that the enhancement
in the $\gamma\gamma$ channel is $\sim 1.3$ (for both $gg$ fusion and VBF) for $\tanb=4$ and $20$ while other final states, in particular $ZZ$,  have close to SM rates. The table also shows that this maximum is achieved for $\sin\alpha\sim 0$.  Thus, $\beta\sim \pi/2$ and $\cos\alpha\sim 1$ yielding SM-like coupling of the $h$ to quarks (see Table~\ref{fermcoups}) and vector bosons. It turns out that in these cases the total enhancement, $\sim 30\%$, 
is provided by the charged Higgs boson loop contribution to the $\gamma\gamma$-coupling. In these same cases, the mass of the heavier Higgs boson is $m_H=225\gev$. 
As such a mass is within the reach of the LHC, it is important to make sure that the $H$ cannot be detected (at least with the current data set). It is easy to see that 
indeed this is the case.
Since $g_{HZZ} \propto \cos(\beta-\alpha)$ and $g_{Hb\anti b,Ht\anti t}\propto \sin\alpha$ one finds that the $H$ decouples from both vector 
bosons and fermions given that  $\alpha\sim 0$ and $\beta\sim \pi/2$.  The $A$ will also be difficult to detect since it has no tree-level $WW,ZZ$ coupling and the $Ab\anti b,At\anti t$ couplings, being proportional to $\cotb$, will be quite suppressed, especially at $\tanb=20$. From Table~\ref{tabhIonly}, we observe that for  $\tanb=4$ and $20$ the corresponding charged Higgs is light,
$\mhpm=90\gev$, \ie\ as small as allowed by LEP2 direct searches in $e^+e^-\to H^+H^-$. Searches for a light $\hpm$ are 
underway at the LHC along the lines described in \cite{Aoki:2011wd}. The most promising $H^\pm$ production and decay process is 
$pp\to t\bar{t}\to H^\pm b W^{\mp} \bar{b} \to \tau \nu b \bar{b} q^\prime \bar{q}$. According to Fig.~3 of \cite{Aoki:2011wd}, 
for the Type I model, the region of $\tanb \lsim 6-7$ for $m_{H^\pm}\sim 90\gev$ could be efficiently explored at the $14\tev$ 
LHC by ATLAS even at the integrated luminosity  of $10\;\text{fb}^{-1}$ -- for more details see \cite{Aoki:2011wd}. 
The existing LHC bounds on $BR(t\to H^+ b)$ obtained assuming $\br(\hpm\to \tau^\pm \nu_\tau)=1$ are only moderately restrictive:  $5\%-1\%$~\cite{Aad:2012tj} 
($4\% -2\%$)\cite{:2012cwb} for masses of the charged Higgs boson $m_{H^\pm} = 90 (80)-160\gev$ in the case of ATLAS (CMS), respectively. These bounds are weakened
in the Type I model where $BR(\hpm\to \tau^\pm\nu_\tau)\simeq 0.7$. Since $BR(t\to H^+ b)\sim 1/\tan^2\beta$, large $\tanb$ suppresses $BR(t\to H^+ b)$. Indeed, it is easy to
verify that for $m_{H^\pm} = 90\gev$ $BR(t\to H^+ b)$ is $\sim 3.8\%$ and $\sim 0.15\%$ for  $\tanb = 4$ and $\tanb=20$, respectively.
So, a charged Higgs yielding enhanced $h\to \gam\gam$ rates in $gg$ fusion and VBF is still completely consistent with current data.

%

\subsection{\bf \boldmath$m_A=125.1\gev$ and $m_h=125\gev$ or $m_H=125\gev$ scenario}

\begin{table}[b]
\begin{tabular}{|c|ccccc|ccc|ccc|c|c|c|}
\hline
$\tan\beta$  &$R^{h+A}_{gg\text{max}}(\gam\gam)$  &$R^h_{gg}(\gam\gam)$  &$R^A_{gg}(\gam\gam)$  &$R^{h+A}_{gg}(ZZ)$  &$R^{h+A}_{gg}(\tau\tau)$  &$R^h_{\text{VBF}}(\gam\gam)$  &$R^h_{\text{VBF}}(ZZ)$  &$R^h_{VH}(b\anti{b})$  &$m_H$  &$m_{H^\pm}$  &$m_{12}$  &$\sin\alpha$  &$\mathcal{A}^h_{H^\pm}/\mathcal{A}$ &$\damu$  \cr  \hline 
 1.2	&1.31	&0.90	&0.41	&1.02	&3.35	&0.83	&0.94	&1.02	&625	&612	& 100	&-0.6	&-0.05	&-3.4 \cr  \hline 
 1.4	&1.21	&0.91	&0.30	&0.99	&2.61	&0.94	&1.03	&0.99	&425	&460	& 100	&-0.6	&-0.05	&-2.9 \cr  \hline 
 1.6	&1.14	&0.91	&0.23	&1.01	&2.32	&0.87	&0.97	&1.01	&425	&360	& 100	&-0.5	&-0.05	&-2.7 \cr  \hline 
 1.8	&1.10	&0.92	&0.18	&1.00	&1.98	&0.94	&1.01	&1.00	&325	&285	& 100	&-0.5	&-0.04	&-2.4 \cr  \hline 
 2.0	&1.08	&0.93	&0.15	&0.98	&1.73	&0.99	&1.04	&0.98	&325	&230	& 100	&-0.5	&-0.04	&-2.2 \cr  \hline 
 3.0	&1.34	&1.29	&0.06	&1.00	&1.31	&1.27	&0.99	&1.00	&225	& 92	& 100	&-0.3	&0.12	&-1.9 \cr  \hline 
 4.0	&1.35	&1.33	&0.03	&0.99	&1.21	&1.24	&0.93	&0.99	&225	& 90	& 100	&-0.1	&0.14	&-1.8 \cr  \hline 
 5.0	&0.96	&0.95	&0.01	&1.00	&1.07	&0.95	&1.00	&1.00	&135	& 90	&  50	&-0.2	&-0.03	&-1.7 \cr  \hline 
 7.0	&1.04	&1.04	&0.01	&0.99	&1.00	&1.06	&1.01	&0.99	&135	& 90	&  50	&-0.2	&0.02	&-1.6 \cr  \hline 
10.0	&0.91	&0.90	&0.01	&0.81	&0.77	&0.99	&0.89	&0.81	&175	&150	&  50	&-0.5	&0.04	&-1.5 \cr  \hline 
15.0	&0.42	&0.42	&0.00	&0.59	&0.67	&0.37	&0.53	&0.59	&225	&250	&  50	& 0.6	&-0.17	&-1.4 \cr  \hline 
20.0	&1.31	&1.31	&0.00	&1.00	&1.00	&1.30	&0.99	&1.00	&225	& 90	&  50	&-0.0	&0.13	&-1.6 \cr  \hline 

\end{tabular}
\vspace{-.1in}
\caption{Table of maximum $R^{h+A}_{gg}(\gam\gam)$ values for the Type I 2HDM with $m_h=125\gev,m_A=125.1\gev$  and  associated $R$ values for other initial and/or final states.  The input parameters that give the maximal $R^{h+A}_{gg}(\gam\gam)$ value are also tabulated. Note that $R(b\anti b)$ values can be obtained from this table by using $R(b\anti b)=R(\tau\tau)$. $\tanb$ values, see Table~\ref{parasum}, for which the full set of constraints cannot be obeyed are omitted}
\label{tabAhI}
\end{table}

%
\begin{figure}[t]
\begin{center}
\includegraphics[width=0.45\textwidth]{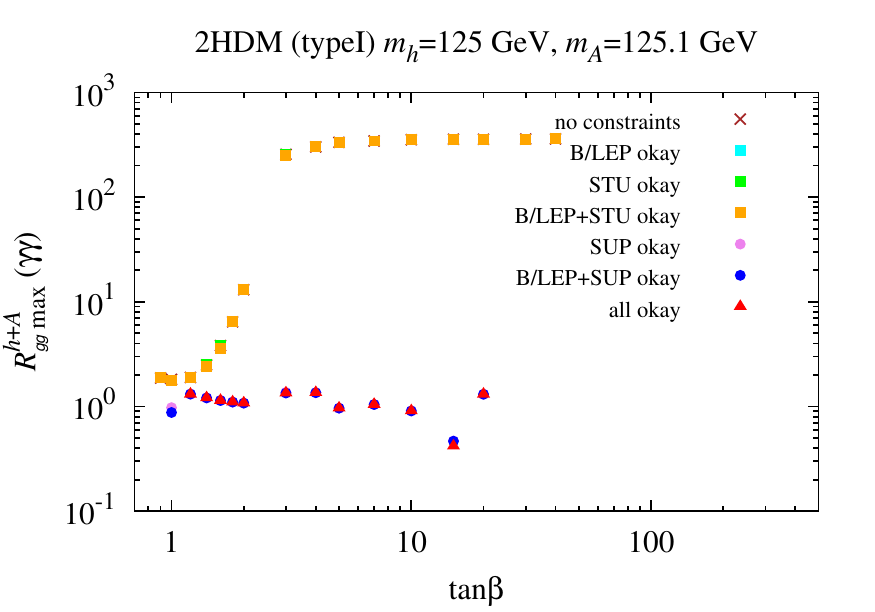}
\includegraphics[width=0.45\textwidth]{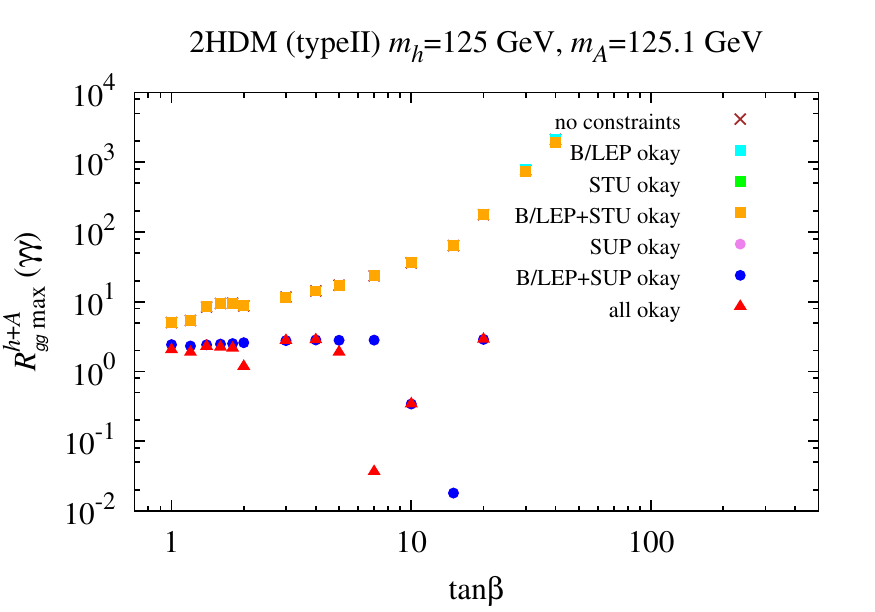}
\includegraphics[width=0.45\textwidth]{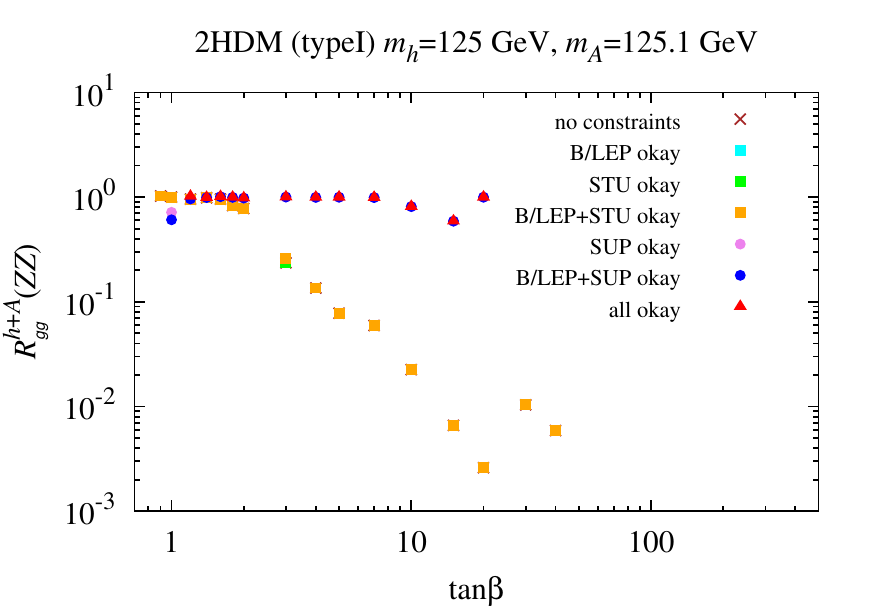}
\includegraphics[width=0.45\textwidth]{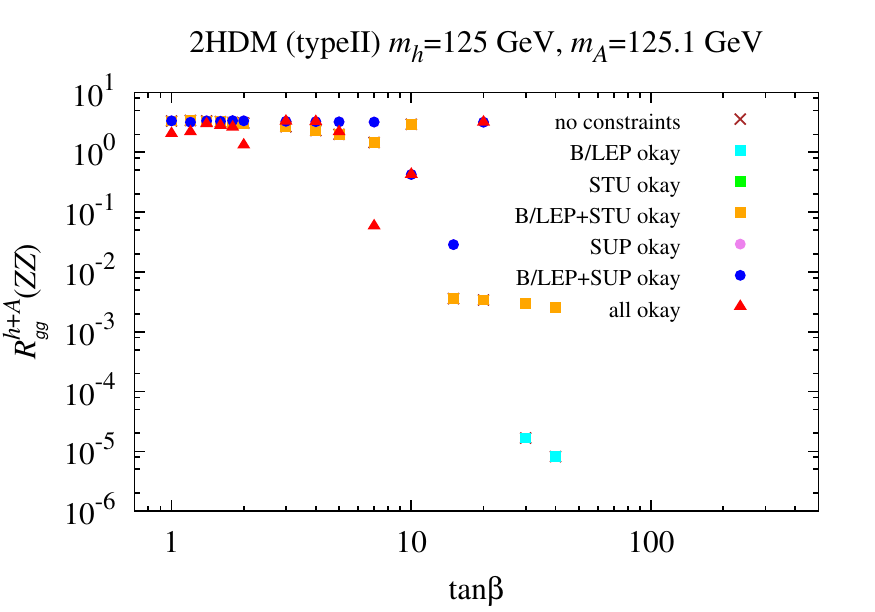}
\includegraphics[width=0.45\textwidth]{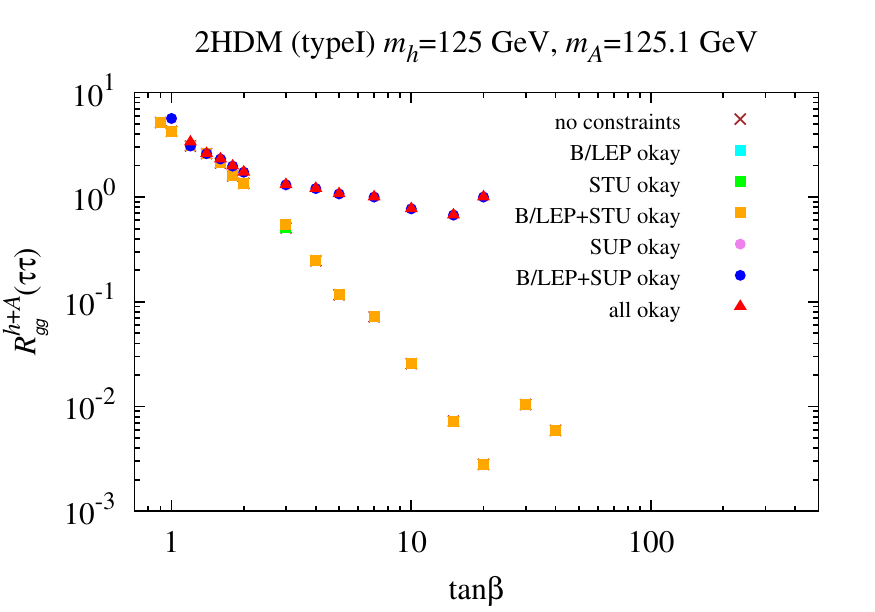}
\includegraphics[width=0.45\textwidth]{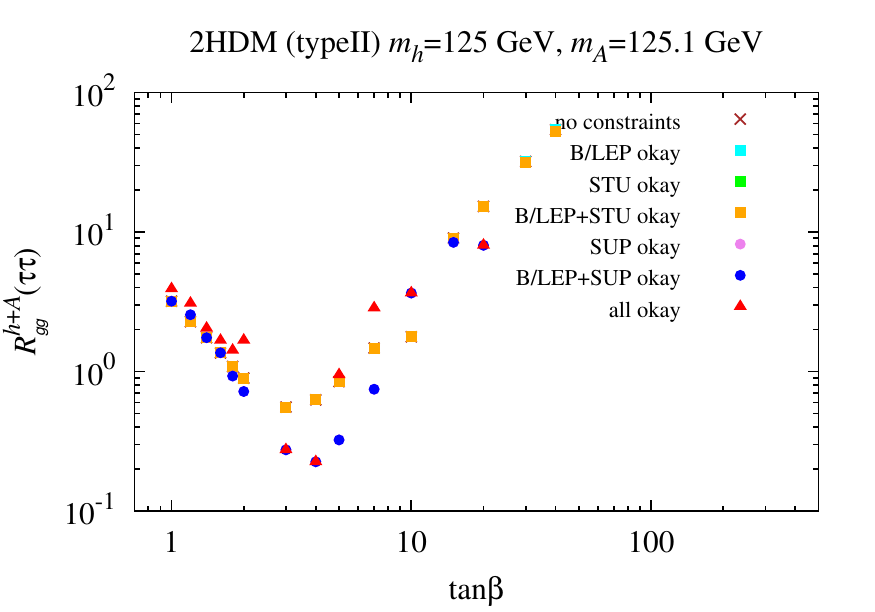}
\end{center}
\vspace{-.25in}
\caption{$R^{h+A}_{gg}(\gam\gam)$ maximum values when $m_h=125\gev,m_A=125.1\gev$ as a function of $\tanb$ after imposing various 
constraints --- see figure legend. { Corresponding $R^h_{gg}(ZZ)$ and $R^h_{gg}(\tau\tau)$ ($=R^h_{gg}(b\anti b)$) are shown in the middle and lower panels.}}
\label{hAonly}
\end{figure}

%
\begin{figure}
\begin{center}
\includegraphics[width=0.45\textwidth]{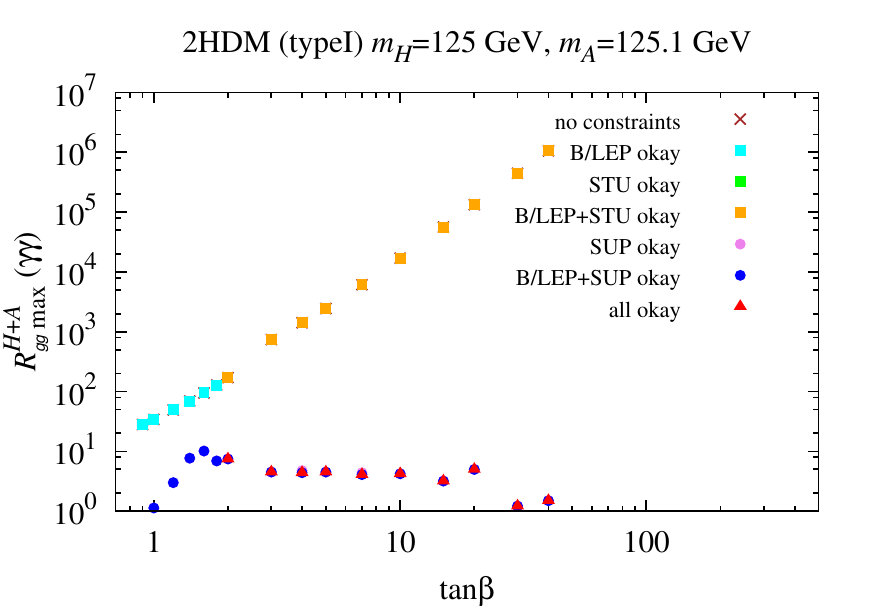}
\includegraphics[width=0.45\textwidth]{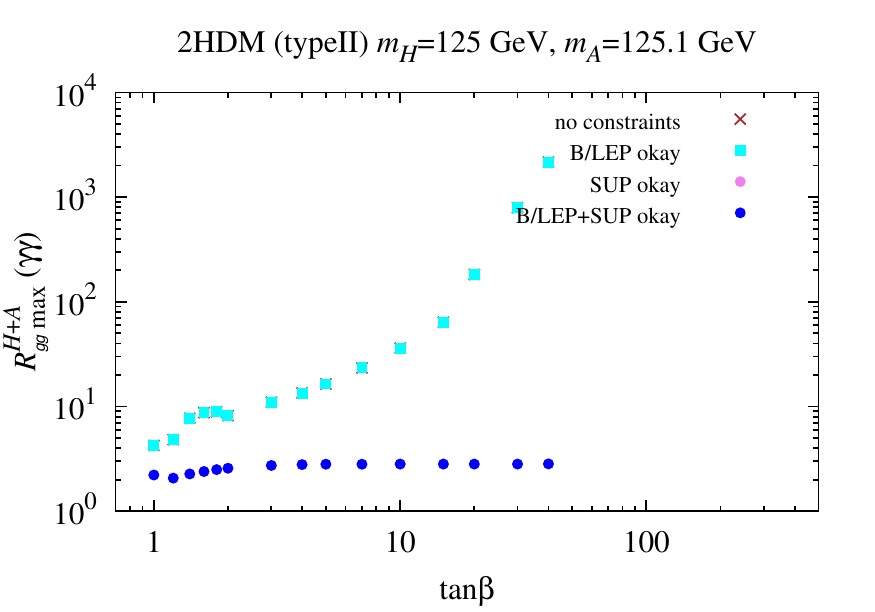}
\includegraphics[width=0.45\textwidth]{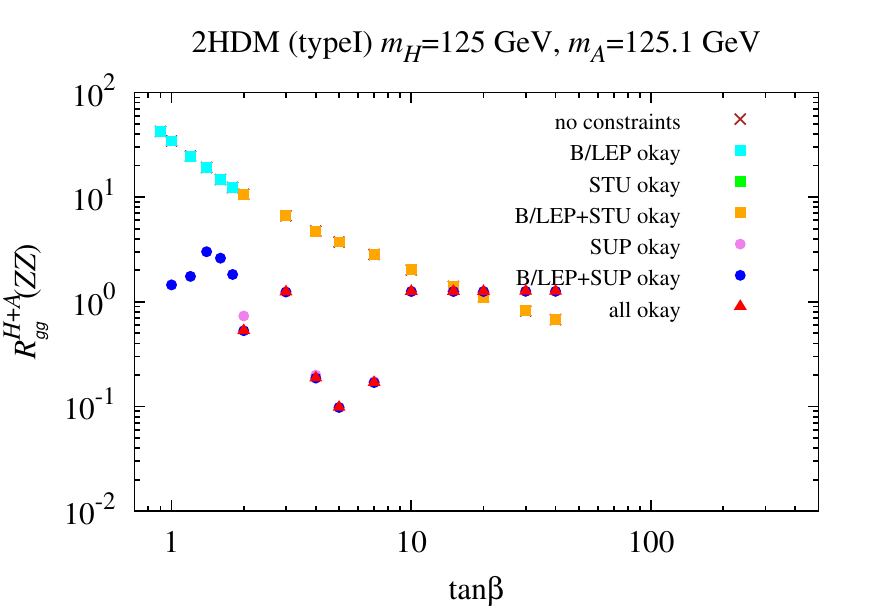}
\includegraphics[width=0.45\textwidth]{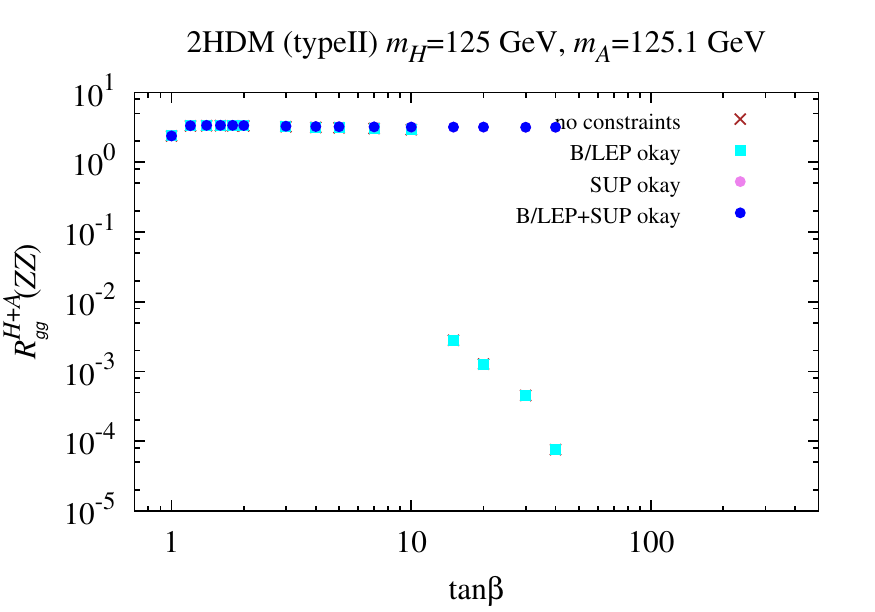}
\includegraphics[width=0.45\textwidth]{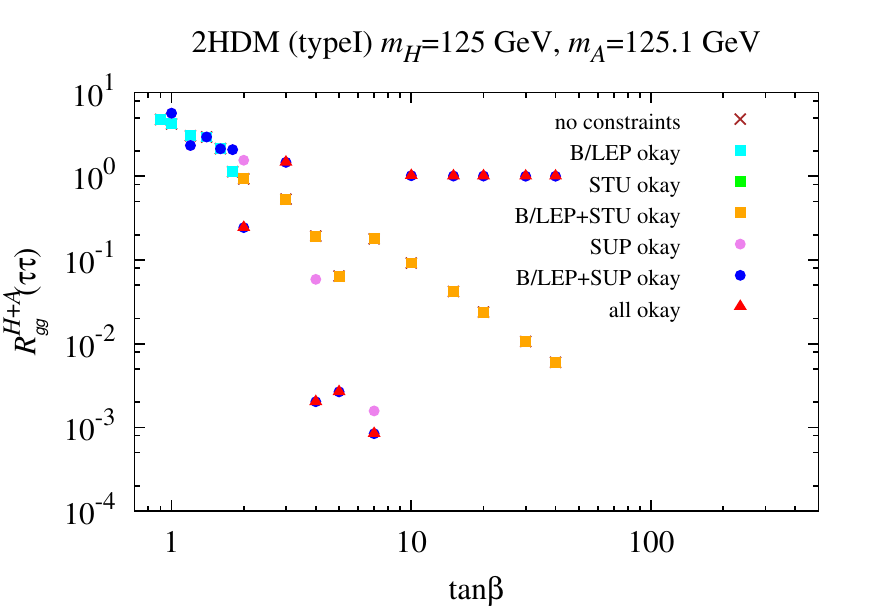}
\includegraphics[width=0.45\textwidth]{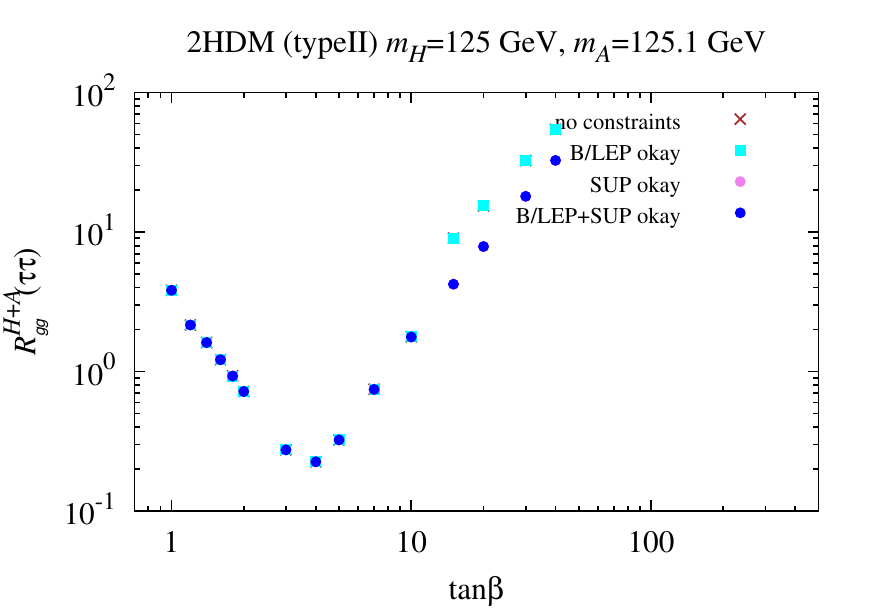}
\end{center}
\vspace{-.25in}
\caption{$R^{h+A}_{gg}(\gam\gam)$ maximum values when $m_H=125\gev,m_A=125.1\gev$ as a function of $\tanb$ after imposing various 
constraints --- see figure legend.  Corresponding $R^h_{gg}(ZZ)$ and $R^h_{gg}(\tau\tau)$ ($=R^h_{gg}(b\anti b)$) are shown in the middle and lower panels.}
\label{HAonly}
\end{figure}
\begin{figure}[t]
\begin{center}
\includegraphics[width=0.45\textwidth,height=0.23\textheight]{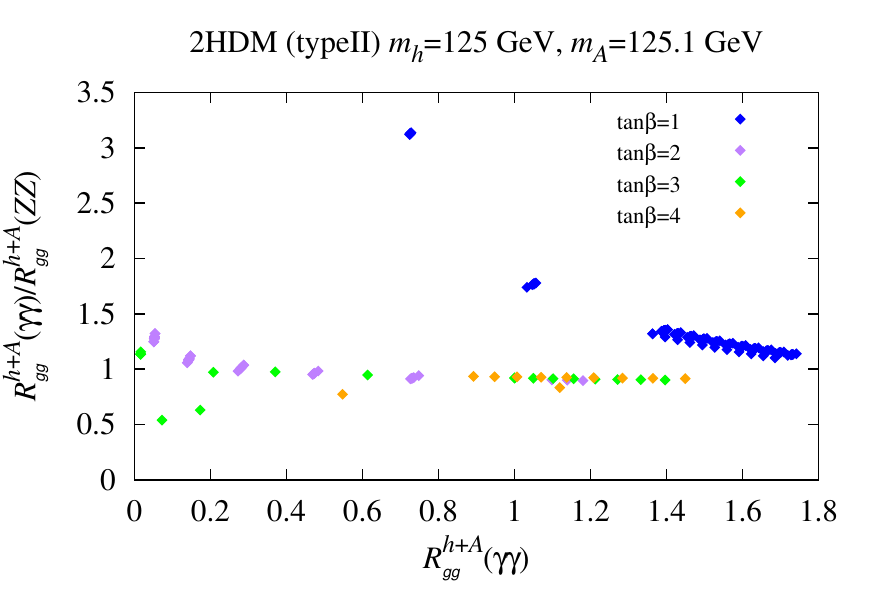}
\includegraphics[width=0.45\textwidth,height=0.23\textheight]{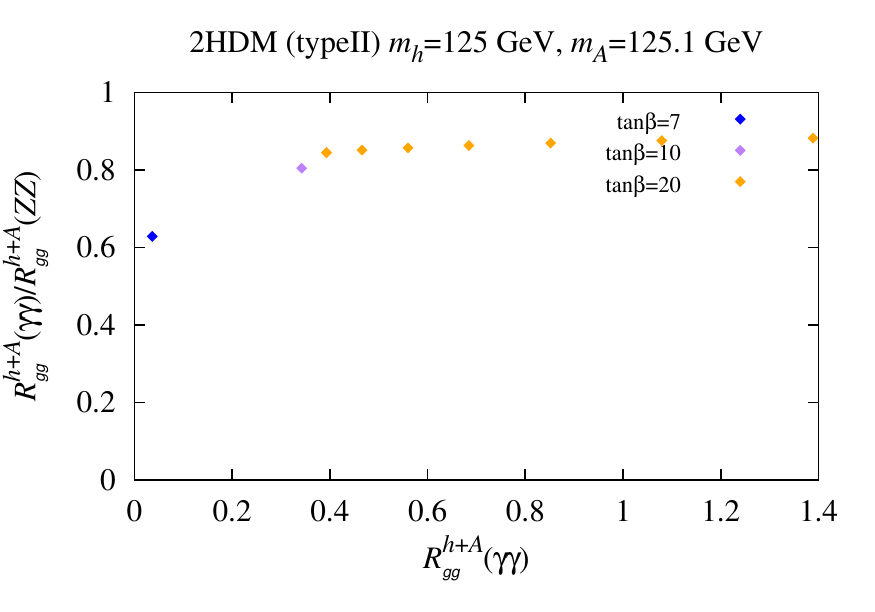}\\
\includegraphics[width=0.45\textwidth,height=0.23\textheight]{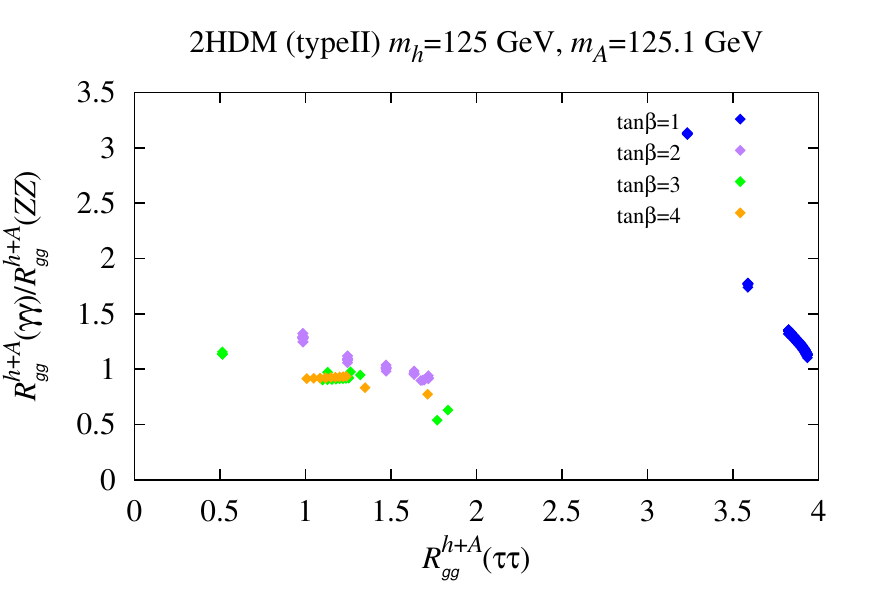}
\includegraphics [width=0.45\textwidth,height=0.23\textheight]{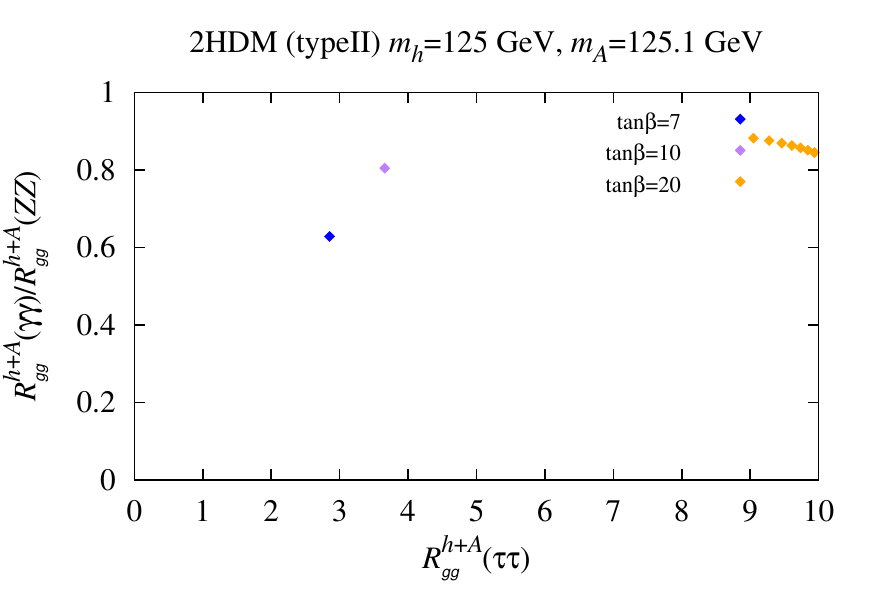}
\end{center}
\vspace{-.25in}
\caption{Correlations between $R_{gg}^{h+A}(\gam\gam)/R^{h+A}_{gg}(ZZ)$ and $R_{gg}^{h+A}(\gam\gam)$ (upper panels)
and  $R^{h+A}_{gg}(\tau\tau)$ (lower panels) for selected values of $\tanb$.}
\label{corbb}
\end{figure}

The signal at $125\gev$ cannot be pure $A$ since the $A$ does not couple to $ZZ$, a final state that is definitely 
present at $125\gev$.  However, one can imagine that the $\cp$-even $h$ or $H$ and the $A$ both have mass close to $125\gev$ 
and that the net $\gam\gam$ rate gets substantial contributions from
both the $h$ or $H$ and the $A$ while only the former contributes to the $ZZ$ rate.  These possibilities are 
explored in Figs.~\ref{hAonly} and \ref{HAonly}, from which we observe that an enhanced $\gam\gam$ rate is only possible for the $m_h=125, m_A=125.1\gev$ choice.
Details for this case appear in Table~\ref{tabAhI}.

For the Type I model, we see from  Table~\ref{tabAhI}  that $R^h_{gg}(\gam\gam)$ is significantly enhanced only for the same $\tanb=4$ and $\tanb=20$
values  as in the case of having (only) $m_h=125\gev$ and that the pseudoscalar contribution  $R^A_{gg}(\gam\gam)$ turns out to be tiny.
However, the contribution to the  $b\anti b$ final state from the $A$ {\it can} be substantial. Given that the top loop dominates both the $Agg$ and $hgg$ coupling one finds  $(C^A_{gg}/C^h_{gg})^2 \sim (3/2)^2 (\cos\beta/\cos\alpha)^2$, where we used  $C^A_{t\anti t}/C^h_{t\anti t}=\cos\beta/\cos\alpha$ from Table~\ref{fermcoups} and the $m_{h,A}\ll 2m_t$ fermionic loop ratio of $A/h=3/2$. As a result, the $A$ can contribute even more to the $b\anti b$ final state rate than the $h$ if $\tanb$ is small. This (unwanted) contribution to the $b\anti b$ final state from $A$ production is apparent from the results for $R^{h+A}_{gg}(b\anti b)$ in Table~\ref{tabAhI} 
for $\tanb=2-4$. In the end, only $\tan\beta=20$  yields both an enhanced $\gam\gam$ rate,  
$R_{gg\, {\rm max}}^{h+A}(\gam\gam)=1.31$, and SM-like rates for the $ZZ$ and $b\anti b$ final states, $R^{h+A}_{gg}(ZZ) = R^{h+A}_{gg}(b\bar{b})=1$. 
For this case $\beta\simeq \pi/2$ and $\alpha=0$ implying that the $h$ couples to fermions and gauge bosons like a SM Higgs boson and 
the enhancement of $R_{gg\, {\rm max}}^{h+A}(\gam\gam)$ is due exclusively to the charged Higgs loop contribution to the $\gam\gam$ couplings. 
\begin{table}[h]
\begin{tabular}{|c|ccccc|ccc|ccc|c|c|c|}
\hline
$\tan\beta$  &$R^{h+A}_{gg\text{max}}(\gam\gam)$  &$R^h_{gg}(\gam\gam)$  &$R^A_{gg}(\gam\gam)$  &$R^{h+A}_{gg}(ZZ)$  &$R^{h+A}_{gg}(\tau\tau)$  &$R^h_{\text{VBF}}(\gam\gam)$  &$R^h_{\text{VBF}}(ZZ)$  &$R^h_{VH}(b\anti{b})$  &$m_H$  &$m_{H^\pm}$  &$m_{12}$  &$\sin\alpha$  &$\mathcal{A}^h_{H^\pm}/\mathcal{A}$ &$\damu$  \cr  \hline 
 1.0	&2.05	&1.58	&0.47	&2.05	&3.91	&0.93	&1.22	&0.65	&525	&500	& 100	&-0.5	&-0.06	& 1.3 \cr  \hline 
 1.2	&1.88	&1.71	&0.17	&2.19	&3.08	&1.05	&1.34	&0.57	&425	&450	& 100	&-0.4	&-0.05	& 1.5 \cr  \hline 
 1.4	&2.29	&2.22	&0.07	&2.99	&2.04	&1.18	&1.59	&0.23	&325	&340	& 100	&-0.2	&-0.05	& 1.9 \cr  \hline 
 1.6	&2.23	&2.20	&0.03	&2.80	&1.67	&1.34	&1.71	&0.28	&225	&320	& 100	&-0.2	&-0.05	& 2.0 \cr  \hline 
 1.8	&2.15	&2.14	&0.01	&2.63	&1.42	&1.44	&1.77	&0.33	&225	&320	& 100	&-0.2	&-0.05	& 2.0 \cr  \hline 
 2.0	&1.18	&1.17	&0.01	&1.31	&1.68	&1.07	&1.20	&0.87	&325	&340	& 100	&-0.4	&-0.05	& 1.5 \cr  \hline 
 3.0	&2.78	&2.78	&0.00	&3.29	&0.27	&2.01	&2.37	&0.00	&225	&320	& 100	&-0.0	&-0.05	& 2.3 \cr  \hline 
 4.0	&2.84	&2.84	&0.00	&3.25	&0.23	&2.24	&2.57	&0.00	&225	&320	& 100	&-0.0	&-0.04	& 2.3 \cr  \hline 
 5.0	&1.89	&1.89	&0.00	&2.19	&0.95	&1.41	&1.64	&0.47	&225	&320	& 100	& 0.1	&-0.05	& 2.7 \cr  \hline 
 7.0	&0.04	&0.04	&0.00	&0.06	&2.85	&0.01	&0.02	&0.75	&325	&320	& 100	& 0.6	&-0.15	& 5.2 \cr  \hline 
10.0	&0.34	&0.34	&0.00	&0.43	&3.66	&0.22	&0.28	&1.23	&325	&320	& 100	& 0.2	&-0.08	& 4.7 \cr  \hline 
20.0	&2.89	&2.89	&0.00	&3.19	&8.03	&2.57	&2.83	&0.00	&225	&320	&  50	&-0.0	&-0.04	& 5.6 \cr  \hline 

\end{tabular}
\vspace{-.1in}
\caption{Table of maximum $R^{h+A}_{gg}(\gam\gam)$ values for the Type II 2HDM with $m_h=125\gev,m_A=125.1\gev$ and associated $R$ values for other initial and/or final states.  The input parameters that give the maximal $R^{h+A}_{gg}(\gam\gam)$ value are also tabulated. Note that $R(b\anti b)$ values can be obtained from this table by using $R(b\anti b)=R(\tau\tau)$. $\tanb$ values. see Table~\ref{parasum}, for which the full set of constraints cannot be obeyed are omitted}
\label{tabAhII}
\end{table}

For the Type II model, see Table~\ref{tabAhII}, the pseudoscalar contribution $R^A_{gg}(\gam\gam)$ 
is also (as for the Type I model) negligible. Thus, the enhancement of $R_{gg}^{h+A}(\gam\gam)$ is essentially the same as that for $R_{gg}^h(\gam\gam)$ for the case when only $m_h=125\gev$, reaching maximum values of order $2-3$.
However, as in the pure $m_h=125\gev$ case, a substantial enhancement  of $R_{gg}^{h+A}(\gam\gam)$
is most often associated with $R_{gg}^{h+A}(ZZ)>R_{gg}^{h+A}(\gam\gam)$ (contrary to the LHC observations).  But this is not {\it always} the case.  Among the $m_h\sim m_A$ scenarios we find  56 points in our parameter space
for which $R^{h+A}_{gg}(ZZ) < 1.3$ and $R_{gg}^{h+A}(\gam\gam) > 1.3$. Unfortunately for all those
points the $\tau\tau$ signal is predicted to be too strong, $R^{h+A}_{gg}(\tau\tau) > 3.82$, a result that is now excluded by the CMS analysis in the gluon fusion dominated $1$-jet trigger mode which finds  $R^{h+A}_{gg}(\tau\tau) < 1.8$ at 95\% CL. This situation
is illustrated in Fig.~\ref{corbb}\,. As  seen from the upper panels in Fig.~\ref{corbb}, 
for $\tanb=1$ there exist points (blue diamonds) such that ${R^h_{gg}(\gam\gam)\ov R^h_{gg}(ZZ) } > 1$ and 
$R_{gg}^{h+A}(\gam\gam) > 1$ (or even $>1.5$). However, the lower 
left panel of Fig.~\ref{corbb} shows that the $R^{h+A}_{gg}(\tau\tau)$ values that correspond to those points are 
greater than $3.5$.

The case with $m_A\sim125\gev$ and $m_H=125\gev$ is less attractive. For the Type I model, 
the constraints are such that once parameters are chosen so that $H$ and $A$ have masses of  $125\gev$ and $125.1\gev$ the maximum 
value achieved for $R_{gg\, {\rm max}}^{H+A}(\gam\gam)$ is rather modest reaching only $1.04$ at small $\tanb$. 
For the Type II model,  as seen in Fig.~\ref{HAonly}, there are no parameter choices for 
which the $H$ and $A$ have a mass of $\sim 125\gev$ while all other constraints are satisfied. 

\subsection{ $m_h=125\gev$ and $m_H=125.1\gev$ scenario}
Finally, we have the case where $m_h=125\gev$ and $m_H=125.1\gev$ and we allow $\mhpm$ and $m_A$ to vary freely.  
Following a similar search strategy, we find that some of the $\tanb$ values previously available when only $m_h=125\gev$ or $m_H=125\gev$ was required are ruled out 
by the full set of constraints and that there is no gain in maximal $R_{gg}^{h+H}(\gam\gam)$ values, and often some loss, relative to 
the cases where only the $h$ or only the $H$ was required to have mass of $125\gev$.\\\\

\begin{figure}
\vspace{-10pt}
\begin{center}
\includegraphics[width=0.45\textwidth,height=0.23\textheight]{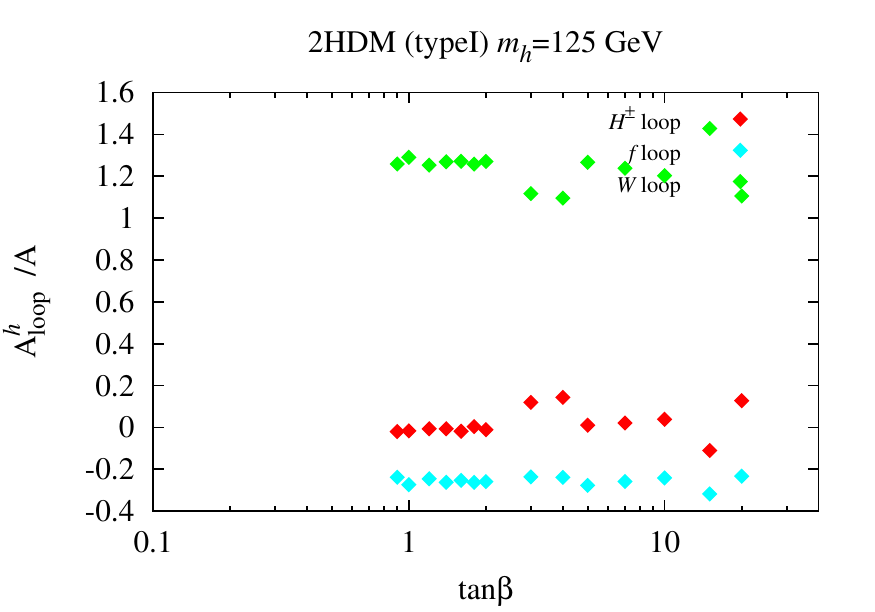}
\includegraphics[width=0.45\textwidth,height=0.23\textheight]{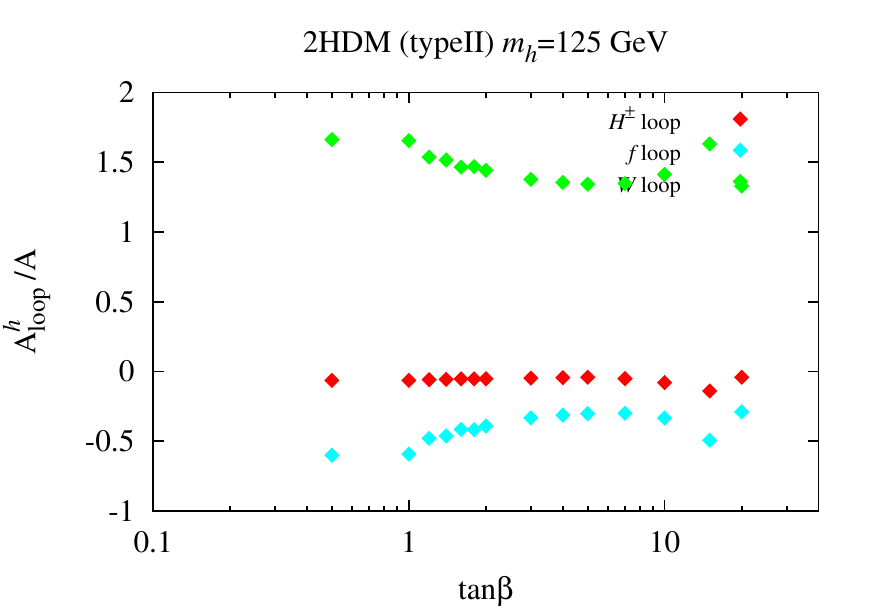}
\includegraphics[width=0.45\textwidth,height=0.23\textheight]{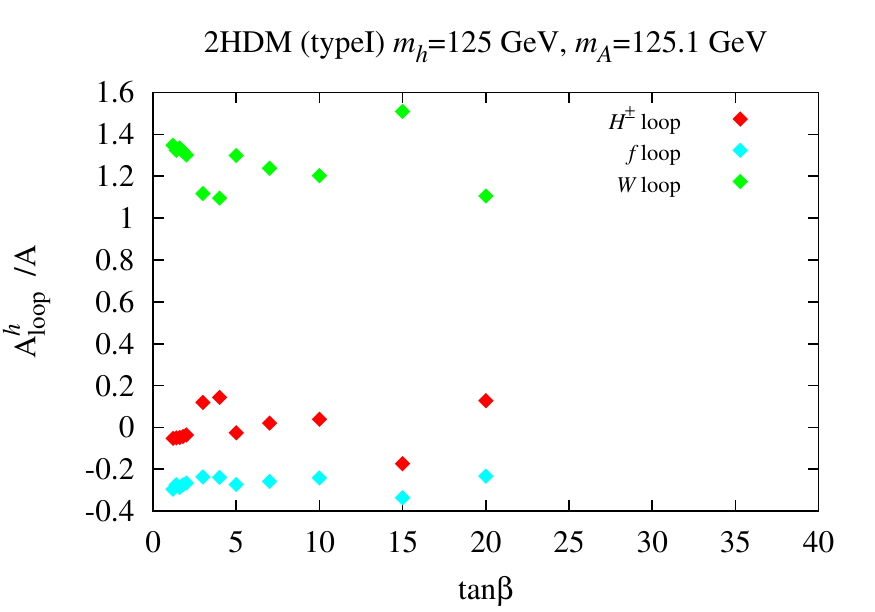}
\includegraphics[width=0.45\textwidth,height=0.23\textheight]{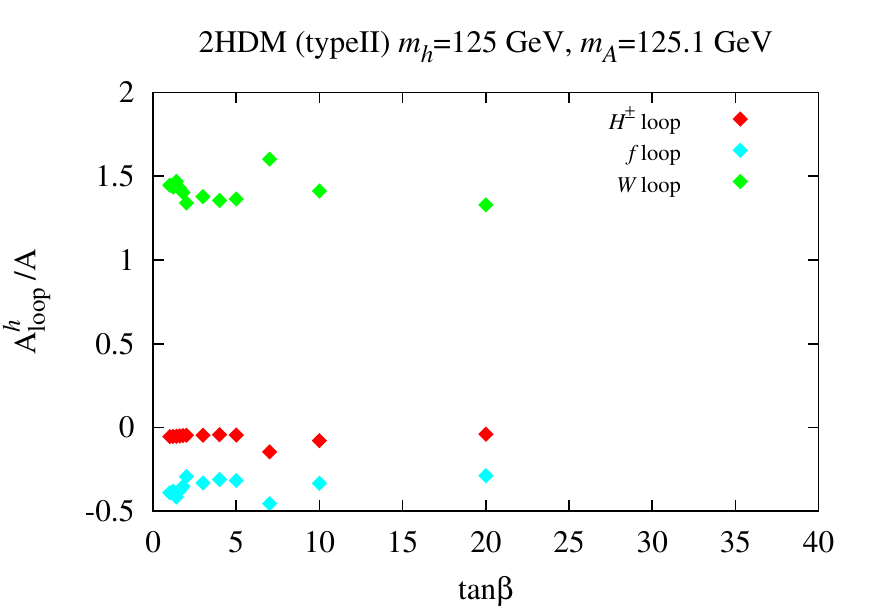}
\end{center}
\vspace{-.25in}
\caption{For the most interesting scenarios we show 
imaginary part of charged Higgs contributions to the $\gamma\gamma$ amplitude normalized to the imaginary 
part of the sum of
all (fermions, $W^+W^-$, $H^+H^-$) contributions as a function of $\tanb$ after imposing all constraints. The 
parameters adopted correspond to maximal $R^{h_i}_{gg}(\gam\gam)$ (or an appropriate sum for degenerate cases).}
\label{HCAmp}
\end{figure}

As discusssed earlier and in \cite{Altmannshofer:2012ar},  the charged Higgs contribution to the $\gam\gam$ coupling loops is sometimes relevant. Therefore,
in Fig.~\ref{HCAmp} we show separately the fermionic loop, $W$ loop and $\hpm$ loop contributions normalized 
to the total amplitude for the most interesting cases of a Type I model  with  $m_h=125\gev$  and with $m_h=125\gev,m_A=125.1\gev$ (left plots). One sees that the $\tanb$ values of 4 and 20 associated with $R^h_{gg}(\gam\gam)\sim 1.3$ are associated with large $A_{\hpm}/A$.  Indeed, in these two cases, the relative charged Higgs contribution reaches nearly $\sim 0.2$ and  is as large as the fermionic contribution, but of the opposite sign. In fact, 
although the dominant loop is the $W$ loop, the  $\hpm$ loop may contribute as much as the dominant (top quark) fermionic loop.  

This should be contrasted with other cases, such as the Type II $m_h=125\gev$ and $m_h=125\gev, m_A=125.1\gev$ cases illustrated in the right-hand plots of Fig.~\ref{HCAmp}.
One finds that the charged Higgs contributions are small when SUP constraints are imposed. 
 In fact, the enhancement of $R^h_{gg}(\gam\gam)$ observed in Fig.~\ref{honly}
prior to imposing SUP is caused just by the charged Higgs loop. When SUP constraints are imposed the charged Higgs amplitude is strongly reduced by the requirement that the quartic couplings not violate the perturbativity condition. 
Note that the SUP constraints can be violated even though all the mass parameters
have been varied within what, a priori, appears to be a reasonable range, namely from a few GeV up to $1000\gev$. This is due to the fact that, for our input, the 
SUP conditions imply a strong constraint on $m_{12}^2$ that comes mainly from the requirement of keeping $\lambda_1$ small enough.

\section{Conclusions}
We have analyzed the Type I and Type II two-Higgs-doublet extensions of the Standard Model with regard to consistency with a significant
enhancement of the gluon-fusion-induced $\gamma\gamma$ signal at the LHC at $\sim 125\gev$, as seen in the ATLAS data set, but possibly not in the CMS results presented at Moriond 2013.
All possible theoretical and experimental constraints have been imposed.
We find that vacuum stability, unitarity and perturbativity play the key role in limiting the maximal possible enhancement which,
in the most interesting scenarios, is generated by the charged Higgs loop contribution to the Higgs to two photon 
decay amplitude.  
Generically, we conclude that the Type II model allows a maximal enhancement of order of $2-3$, whereas within
the Type I model the maximal enhancement is limited to  $\lsim 1.3$. 
Moriond 2013 ATLAS results suggest an enhancement for $gg\to h\to \gam\gam$ of order $1.6$ (but with large errors). Only Type II models can give such a large value.

However,  we find  that in the Type II model the parameters that give $R^h_{gg}(\gam\gam)\sim 1.6$ are characterized by $R^h_{gg}(ZZ)\sim (3/2) R^h_{gg}(\gam\gam)$, a result that is  inconsistent with the ATLAS central value of $R^h_{gg}(ZZ) \sim 1.5$.   Thus, the Type II model cannot describe the ATLAS data if only the $h$ resides at $125\gev$.  
Similar statements apply to the case of the heavier $H$ having a mass of $125\gev$. In contrast, the CMS data suggests values of $R^h_{gg}(\gam\gam)<1$ and  $R^h_{gg}(ZZ)\sim 1$, easily obtained in the Type II model context.
Next, we considered Type II models  with approximately degenerate Higgs bosons at $125\gev$.  We found 
that for $1\leq \tanb\leq 5 $ there exist theoretically consistent parameter choices for Type II models for which $R^{h+A}_{gg}(\gam\gam) \sim R^{h+A}_{gg}(ZZ) \sim 1.6$, fully consistent with the ATLAS results.
Unfortunately, 
in these cases $R^{h+A}_{gg}(\tau\tau) >3.75$, a value far above that observed. Thus, the Type II 2HDMs cannot yield  
$R^{h+A}_{gg}(\gam\gam)\sim 1.6$ without conflicting with other observables.  In short, the Type II model is unable to give a significantly enhanced $gg\to h \to \gam\gam$ signal while maintaining consistency with other channels. 

In the case of the Type I model, the maximal $R^h_{gg}(\gam\gam)$ is  of  order of $1.3$, as found if $\tanb=4$ or  $20$. In these cases, 
$R^h_{gg}(ZZ)$ and $R^h_{gg}(\tau\tau)$ are of order 1.  For these scenarios, the charged Higgs
is light, $m_{H^\pm}=90\gev$.  (Despite this small mass, there is no conflict with LHC data due to the fact that $BR(t\to H^+b)\sim 1/\tan^2\beta$ is small enough to be below current limits.)  Thus, Type I models could provide a consistent picture if the LHC results converge to only a modest enhancement for $R^h_{gg}(\gam\gam)\lsim 1.3$.  

Overall, if $R^h_{gg}(\gam\gam)$ is definitively measured to have a value much above 1.3 while the $ZZ$ and/or $\tau\tau$ channels show little enhancement then there is no consistent 2HDM description. One must go beyond the 2HDM to include new physics such as supersymmetry.

\section*{Acknowledgments} 
This work has been supported in part by US DOE grant DE-FG03-91ER40674 and by the National Science Centre (Poland) as a research project, decision no DEC-2011/01/B/ST2/00438. This work was supported by the Foundation for Polish Science International PhD Projects Programme co-financed by the EU European Regional Development Fund. JFG thanks the Galileo Galilei Institute and the LPSC Grenoble 2013 Higgs Workshop for support  during the completion of this work.

\end{document}